\documentclass[review]{elsarticle}
\usepackage{pifont}
\usepackage{enumitem}
\renewcommand{\figurename}{Fig.}
\usepackage{subfigure}
\usepackage{amsthm}
\usepackage{gensymb}
\usepackage[hyphens]{url}
\usepackage[top=2 cm, left=1.15in, bottom=2 cm, right=1.15in]{geometry} 
\theoremstyle{definition}

\theoremstyle{remark}

\usepackage[ruled]{algorithm2e}
\usepackage{algorithmic}
\usepackage{mathtools}
\usepackage{amsthm}
\usepackage{float}
\usepackage{xcolor}
\usepackage{amsfonts}
\usepackage{amssymb}
\renewcommand{\figurename}{Fig.}
\usepackage{mathrsfs}
\usepackage{graphicx}
\usepackage{subfigure}
\usepackage{amsthm}
\usepackage{amsmath}
\usepackage{tabularx}
\usepackage{multirow}
\usepackage{adjustbox}
\usepackage{tablefootnote}
\usepackage{threeparttable}
\usepackage{blindtext}
\usepackage{amsmath, amsthm, amsfonts}
\usepackage[english]{babel}
\usepackage{tabularx}
\usepackage{mathtools}
\usepackage[ruled]{algorithm2e}
\usepackage{algorithmic}
\usepackage{tikz}
\usepackage{caption}
\usepackage{algorithmic}
\usepackage{babel}
\usepackage{array}
\usepackage{xcolor,colortbl}
\usepackage{lettrine}
\usepackage{pdfpages}
\usepackage{setspace}
\usepackage{multicol}
\usepackage{subfigure}
\usepackage{fancyvrb}
\usepackage{wrapfig}
\usepackage{caption}
\usepackage{subcaption}
\usepackage[normalem]{ulem} 

\makeatletter
\def\ps@pprintTitle{%
  \let\@oddhead\@empty
  \let\@evenhead\@empty
  \let\@oddfoot\@empty
  \let\@evenfoot\@oddfoot
}
\makeatother
\title{A review on Machine Learning based User-Centric Multimedia Streaming Techniques}
\author{Monalisa Ghosh$^{1}$ and Chetna Singhal$^{\dag,1}$}
\ead{monalisa11@iitkgp.ac.in; chetna.iitd@gmail.com}
\address{
$^1$Indian Institute of Technology, Kharagpur, INDIA\\
$^\dag$INRIA, France}
\usepackage{lineno}
\begin{document}
\begin{frontmatter}
\begin{abstract} 
The multimedia content and streaming are a major means of information exchange in the modern era and there is an increasing demand for such services. This coupled with the advancement of future wireless networks B5G/6G and the proliferation of intelligent handheld mobile devices, has facilitated the availability of multimedia content to heterogeneous mobile users. Apart from the conventional video, the 360$\degree$ videos have gained significant attention and are quickly emerging as the popular multimedia format for virtual reality experiences. All formats of videos (conventional and 360$\degree$) undergo processing, compression, and transmission across dynamic wireless channels with restricted bandwidth to facilitate the streaming services. This causes video impairments, leading to quality degradation and poses challenges for the content providers in delivering good Quality-of-Experience (QoE) to the viewers. The QoE is a prominent subjective measure of quality, which has become a crucial component in assessing multimedia services and operations. So, there has been a growing preference for QoE-aware multimedia services over heterogeneous networks with a need to address design issues like how to evaluate and quantify end-to-end QoE. Efficient multimedia streaming techniques can improve the service quality while dealing with dynamic network and end-user challenges. A paradigm shift in user-centric multimedia services is envisioned with a focus on Machine Learning (ML) based QoE modeling and streaming strategies. This survey paper presents a comprehensive overview of the overall and continuous, time varying QoE modeling for the purpose of QoE management in multimedia services. It also examines the recent research on intelligent and adaptive multimedia streaming strategies, with a special emphasis on ML based techniques for video (conventional and 360$\degree$) streaming. This paper discusses the overall and continuous QoE modeling to optimize the end-user viewing experience, efficient video streaming with a focus on user-centric  strategies, associated datasets for modeling and streaming, along with existing shortcoming and open challenges.
\end{abstract}


\end{frontmatter}

\section{Introduction}\label{Intro}
The popularity of video data and video streaming services has increased exponentially in the recent years. Numerous multimedia applications that include video teleconferencing,  video streaming and video-on-demand will dominate the next-generation wireless networks due to the widespread use of internet. Besides the use of conventional video format, 360$\degree$ videos are being used for the augmented reality (AR) and virtual reality (VR) applications that are gaining immense popularity. The emergence of the Metaverse has led to the proliferation of AR and VR in the field of entertainment. In such applications the 360$\degree$ videos are the preferred multimedia format that provide immersive viewing experience to the end-users.
 \par Video delivery has emerged as a major component in the bulk of the present day's internet traffic (more than 70\%) and is expected to grow even further.  According to Ericsson reports \cite{ericsson_mobility}, globally the average data consumption per smart phone is expected to exceed 21 GB per month by the end of 2024. Particularly, video devices will raise the volume of existing traffic, accounting for 74\% of total traffic by the end of 2024.  On a global scale, the average monthly consumption of video streaming will amount to 16.3 GB. 
 The UltraHigh Definition (UHD), or 4K, and UHD-2 (8K) video streaming has increased the impact of video devices on traffic because the bit-rate of 4K video, which is 15 to 18 Mbps, and the bit-rate of 8K, which is 20 to 26 Mbps, is more than twice/thrice that of High Definition (HD) and nine times that of Standard Definition (SD) videos. Sixty-six percent (66\%) of all new flat-panel TV sets are predicted to be UHD. 
 As per the latest Ericsson Mobility Report \cite{ericsson_mobility}, it is estimated that 5G will account for nearly 75\% of mobile data traffic by 2029.  Furthermore, the number of 5G mobile subscriptions is expected to reach a total of 5.6 billion in 2029. Mobile video transmission is undoubtedly a crucial service provided by the 5G mobile networks, and will continue to be so in the upcoming beyond 5G (B5G) era. Service providers are under immense pressure to enhance Quality of Service (QoS) due to the exponential growth in wireless data utilization (primarily multimedia). Increasing video traffic over wireless networks increases the demand for superior multimedia content delivery.  
\par There is an increasing demand for multimedia applications. Video content streaming, exchange and sharing of video-based information are becoming more and more popular among a large number of subscribers. This has compelled service providers to deliver video content of superior quality. Thus, there is an increasing need to ensure that users enjoy a higher Quality of Experience (QoE). 
In our day-to-day lives, we rely heavily on video content sharing through video calling apps and social media content uploads (in Facebook, Instagram).
The majority of popular online services stream video to heterogeneous devices over bandwidth constrained communication networks.
\par The limited bandwidth and unreliable channel conditions pose further challenge.
In order to meet the requirements of the users, high data-rate transmission is required. 
To facilitate the transmission of HD videos at an increased data rate, it is imperative to employ a system that possesses the capability to adapt its configuration based on the prevailing network conditions. 
A feasible solution is the presence of feedback mechanism between the server and the client, which enables the regular updating of the channel state matrix and other user-centric feedback factors at the server. Recently, DASH \cite{edqd, adapt360}—Dynamic Adaptive Streaming over HTTP— has become the industry standard for online video streaming. For DASH systems, a variety of rate adaption techniques are suggested to provide video quality in accordance with the network throughput.
\par In AR/VR applications, the seamless streaming of 360$\degree$ video content there is a need for greater bandwidth and reduced latency compared to existing methods of 2D video delivery. Meeting the bandwidth needs becomes increasingly challenging when streaming the same content to several VR clients. 
A conventional setup for viewing 360$\degree$  video involves a user engaging with the scene via a Head-Mounted Display (HMD) device such as the Oculus Rift, Samsung Gear VR, HTC VIVE, Google Cardboard, and Daydream. The 360$\degree$ video adaptation can be done in a manner similar to conventional videos, as in DASH. A number of research solutions that aid in facilitating an immersive visual experience of 360$\degree$ video include viewport-dependent/independent and tile-based adaptive streaming. Viewport independent solutions often lead to wastage of network resources as entire content is streamed to the users irrespective of their viewport position.
\subsection{Survey novelty and contributions}
This article provides a state-of-the-art survey on the existing solutions for predicting video quality and the methods and strategies that can be used to improve the streaming experience. The major contributions of this survey are as follows:
\begin{enumerate}
\item To provide a review of different existing and emerging video quality assessment models that forecast the video QoE over short duration. 
\vspace{-2 mm}
\item To provide an overview of continuous, time varying video QoE estimation models that are particularly useful in streaming sessions.
\vspace{-2 mm}
\item Survey on efficient user-centric multimedia streaming techniques that have been developed so far for DASH-based streaming, and 360$\degree$ immersive streaming applications. 
\vspace{-2 mm}
\item A thorough analysis of the most recent Machine Learning (ML)-based QoE prediction models and adaptive streaming techniques.
\vspace{-2 mm}
\item  To provide a list of open source datasets that are accessible  publicly for QoE modeling and adaptive streaming of 2D and 360$\degree$ videos.
\vspace{-2 mm}
\item The summary of findings, list of open challenges and future scope are highlighted.
\end{enumerate}
This review provides an analysis of the recent and developing techniques for user-centric multimedia streaming and a thorough examination of the latest ML-based models for predicting QoE. The structure of the survey is as follows: Section \ref{s:need_for_qoe} starts with a discussion on QoE definition, assessment methodologies, need for QoE modeling and correlation existing between Video Quality Assessment (VQA) measures and subjective video quality. In Section \ref{s:user_centric}, we briefly introduce the QoE driven, intelligent and adaptive multimedia streaming. In Section \ref{s:evaluation}, we present evaluation metrics and key considerations for ML-based multimedia streaming. In Section \ref{s:mul_stream}, we discuss the various state-of-the-art models for overall and continuous, time-varying QoE prediction. Section \ref{s:iavs} presents an elaborate review of the intelligent and adaptive streaming techniques for 2D and 360$\degree$ videos. Section \ref{s:data_per} contains several video quality databases and performance metrics that can serve as a helpful tool for future researchers in designing and validating the models. These include the subjective datasets, network traces, 360$\degree$ viewport traces, head movement, and eye tracking datasets as well as different performance metrics to evaluate the effectiveness of the QoE prediction models. In Section \ref{s:open_challenge}, we identify the issues/ challenges that need to be addressed in the future research. We provide a concise summary of our observations and conclude the paper in Section \ref{s:con}.
\section{Need for video QoE Modeling in user-centric multimedia streaming}
\label{s:need_for_qoe}
\subsection{{{Video QoE: Definition and assessment methodologies}}}
\begin{figure}[!htp]
	\centering
	\includegraphics[width=3.4in]{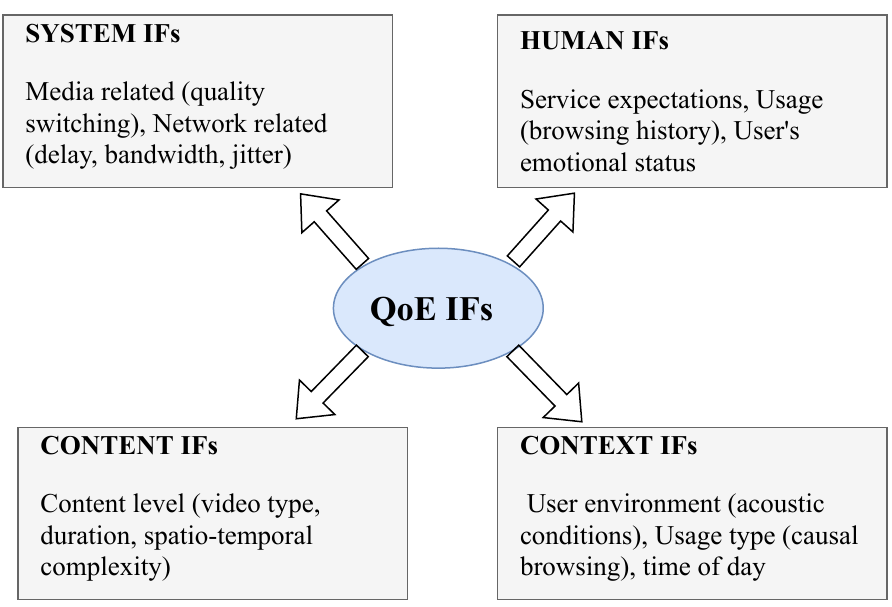}
	\caption{{Factors influencing QoE}}
 \label{f:qoeif}
\end{figure}
According to the European Union (EU) Qualinet Community \cite{eu}, QoE is defined as \textit{``the degree of delight or annoyance of the user of an application or service. It results from the fulfillment of his/her expectations with respect to the utility and/enjoyment of the application or service in the light of the user's personality and current state"}. As per International Telecommunications Union (ITU), \textit{``QoE is defined as the overall quality of an application or a service as perceived subjectively by the end user"} \cite{qoedefinition}. Several factors influencing QoE include- System Influence Factors (IFs), Human IFs, Context IFs, and Content IFs. Fig. \ref{f:qoeif} shows different factors influencing QoE. System IFs are largely concerned with technical elements of quality. Human IFs include, but are not limited to, a user's physical and mental condition, state of mind, memory, and focus, as well as their needs from the service, recency effects, prior use of the application, and more.
Factors like location, end-user surrounding, period of the day, merely casual surfing, service intake period (off-hours, peak hours), are considered as context IFs.
The content IFs focus on the content's individual traits that include content-related details about the service or application under examination. 
\begin{table}[!t]
    \centering
    \caption{Video quality and corresponding opinion scores of users.}
    \label{tab:opinion}
    \begin{tabular}{|c|c|}
     \hline
     \textbf{ Video quality}   &  \textbf{Opinion score}\\ \hline
     Excellent     &  5\\ \hline
     Good & 4 \\  \hline
     Fair & 3\\  \hline
     Poor & 2 \\  \hline
     Bad & 1  \\ \hline
    \end{tabular}
\end{table}

 \begin{table*}[!t]
   \footnotesize
 \centering
    \caption{Features: Definitions and Equations for VQA/IQA metrics.} 
   \label{t:acronym}
\footnotesize
   \begin{tabular}{|p{10pt}|p{60pt}|p{138pt}|p{200pt}|}
  \hline\!\!{\!VQA}& { Definition} & {Equations} & {Variables} \\      \hline 
  FR & Peak Signal to Noise Ratio (PSNR) & \vspace{0.005 cm} 
\shortstack{\scalebox{0.8}{$
     PSNR=20\log_{10}(255/\sqrt{MSE})
   $}\\
    \scalebox{0.8}{$
     \textit{MSE}=  \sum\limits_{i=1}^{\aleph_{1}}\sum\limits_{j=1}^{\aleph_{2}}{\left[X(i,j)-Y(i,j)\right]}^2/(\aleph_{1}\cdot\aleph_{2})
$}}
      & $X(i,j)$ is the  reference frame of original video, $Y(i,j)$ is the reference frame of distorted video; $\aleph_{1}.\aleph_{2}$ is video resolution\\       \hline  
    FR & Non-Content-based
Perceptual PSNR (NC-PSNR) \cite{ncp_psnr} &\vspace{0.005 cm}\!\!\!\!\shortstack{{\scalebox{0.7}{\textit{NC-PSNR}=}}\scalebox{0.85}{$
      10\,\,log \frac{{I}_{max}^{2}}{\sum_{i,j}^{}(I_x(i,j)-{I}_{y}(i,j))^{2}.\tilde{w}(i,j)} \,\,
      $} \\ 
      \scalebox{0.8}{$
      w(i,j)=\underset{(\acute{i},\acute{j})\in \textbf{V}_{i,j}}{\text{max}}v(\acute{i},\acute{j})\,\,
      $}}
      &  {$v(i,j)=u\Big{(}-360\big{(}\frac{i-1}{\aleph_1-1}-\frac{1}{2}\big{)},-180\big{(}\frac{j-1}{\aleph_2-1}-\frac{1}{2}\big{)}\Big{)}$} is pixel alignment likelihood in viewing direction.  $I_x(i,j), I_y(i,j)$ are pixel intensities in original and distorted video, resp. $\tilde{w}(i,j)$ is normalized non-content-based weight map, $w(i,j)$. $\textbf{V}$ is viewport viewing directions.\\ 
      \hline   FR &   Weighted Craster Parabolic Projection PSNR (WCPSNR) \cite{psnr_vr} & 
\vspace{0.005 cm} \shortstack{
\scalebox{0.8}{
$
WCPSNR=10 log_{10} \dfrac{255^{2}}{WMSE} \,\,
$} \\
      \!\!\!{\scalebox{0.7}{\textit{WMSE}=}}\scalebox{0.85}{$
      \frac{\sum_{j=0}^{\aleph_{2}-1}\sum_{i=0}^{\aleph_{1}-1}((a_{i},b_{i})_{x}-(a_{i},b_{i})_{y})^{2}J_{j}}{\sum_{j=0}^{\aleph_{2}-1}\sum_{i=0}^{\aleph_{1}-1}J_{j}} \,\,
  $} }
      & 
    $(a_{i},b_{j})_{x}$, $(a_{i},b_{j})_{y}$ denote nearest-neighbor pixel intensities at sampling points in the projected frame following craster parabolic transformation of original and distorted frames, resp. $J$ is Jacobian determinant for the transformation between cartesian and spherical coordinates.\\   
      \hline FR &  
          Structural Similarity Index Measurement (SSIM) \cite{ssim} &  \vspace{0.005cm}\scalebox{0.8}{$
          SSIM=\dfrac{(2\mu_X\mu_Y\!+\!\mathsf{C}_1)(2\sigma_{XY}\!+\!\mathsf{C}_2)}{({\mu_X}^2\!+\!{\mu_Y}^2\!+\!\mathsf{C}_1)({\sigma_X}^2\!+\!{\sigma_Y}^2\!+\!\mathsf{C}_2)}
$
          } 
          & $\mathsf{C}_1$,  $\mathsf{C}_2$ are constants;  $\mu_X$, $\mu_Y$ and $\sigma_X$, $\sigma_Y$ are mean and standard deviation of the original and distorted frame luminance intensities, resp.\\ 
      \hline  FR &Multi-Scale SSIM (MS-SSIM) & \vspace{0.001cm}
      \scalebox{0.7}{\textit{MS-SSIM}=} \scalebox{0.8}{
  $
     {[l_U(X,Y)]}^{\alpha_{U}} 
      \prod\limits_{j=1}^{U}[c_j{\left (\textit{X,Y}\right )}]^{\beta_{j} }{\left [s_j{\left (X,Y\right )}^{\gamma_{j} }\right ]} \,\,
    $
      }
      & $ \alpha_{U}$, $ \beta_{u}$, and  $\gamma_{u}$ are relative significance parameters of luminance $l_U(X,Y)$, contrast $c(X,Y)$ and structure $s(X,Y)$ components\\      
      \hline {FR} &  
          Weighted-to-Spherically Uniform SSIM (WS-SSIM) \cite{ssim_weighted} &  \vspace{0.001cm}\scalebox{0.7}{\textit{WS-SSIM\!=}}
          \scalebox{0.8}{$
          \dfrac{SSIM\times w_{ck}}{\sum\limits_{k=1}^{N}w_{ck}} \,\,
$} 
          & $w(i,j)=cos((j-\frac{\aleph_{1}}{2}+\frac{1}{2}) \times \frac{\pi}{2})$, $(i,j)$ is current pixel position on projected plane. $w_{ck}$ is weight of center point of $k^{th}$ sliding window, $N$ is number of slides.\\ 
\hline  FR
       & M-Singular Value Decomposition (M-SVD) \cite{d61shnaysvd} &  \vspace{0.001cm}\shortstack{
        \scalebox{0.7}{
  $
       M-SVD=\!{\sum\limits_{i=1}^{v \times v}{|D_{i}\! -\!D_{mi}|}}/(v.v) \,\,
    $}  \\
       \scalebox{0.7}{$
       D_{i}=\sqrt{\sum\limits_{i=1}^{\kappa_{b}}{\left( x_i -{y_i}\right)}^2} \,\,
$}}
       & $v\!=\!\aleph_{1}.\aleph_{2}/\kappa_{b} $; $ \kappa_{b} $ denotes the block size.\hspace{0.1cm} $ D_{mi} $ is the mid-point of sorted $ D_{i}$; \hspace{0.1cm} $ x_i $ and ${y_i}$ denotes  the original and distorted block, resp. \\     
      \hline   RR & Space-Time Generalized Entropic Differences (ST-GREED) \cite{stgreed} & \vspace{0.01cm}\shortstack{
        \scalebox{0.8}{$
        GREED = f(SG,TG_l) \,\,
$} \\
        \scalebox{0.8}{$
        SG=\frac{1}{T}\sum\limits_{t=1}^{T}\big{(}\frac{1}{P}\sum\limits_{p=1}^{P}\left|\phi_{pt}^{Y}-\phi_{pt}^{X} \right| \big{)} \,\,
$}\\
        \scalebox{0.7}{\!\!\!\!$
        TG_{l}\!=\!\frac{1}{T}\sum\limits_{t=1}^{T}\Big{(}\frac{1}{P}\sum\limits_{p=1}^{P}|(1+| \varepsilon _{lpt}^{Y}-\varepsilon_{lpt}^{\acute{X}} |) \frac{\varepsilon_{lpt}^{X}+1}{\varepsilon_{lpt}^{\acute{X}}+1}-1 | \Big{)}
$}}
      & $f(.)$ is mapping function; $\phi_{pt}^{Y}$ and  $\phi_{pt}^{X}$ are modified entropies of distorted and original frames, resp.; $p$ is frame patches; $l$ is sub-band; $t$- istime instant; $\varepsilon_{lpt}^{X}, \varepsilon_{lpt}^{\acute{X}}$, and $\varepsilon_{lpt}^{Y}$ are scaled entropies of original, distorted, and pseudo-reference frames, resp.\\
      \hline  {RR} & Spatial Efficient Entropic Differencing (SpEED) \cite{speedqa}& \vspace{0.01cm}\scalebox{0.9}{$    SpEED = \frac{1}{B_k}\sum\limits_{b=1}^{B_k} \left|h_{bkX}-h_{bkY} \right| \,\,
$}
      & $h_{bkX}$ and $h_{bkY}$ are locally weighted entropies of original and distorted luminance frames, resp. in block $b$ and scale $k$. $B_k$ is scalar quantity per block, based on original and distorted frames. \\ 
 \hline  {NR} & Visual Perception Natural Image Quality Evaluator (VP-NIQUE) \cite{vpniqe} & \vspace{0.01cm}\shortstack{\scalebox{0.9}{$
Q_{VP-NIQE}=\frac{1}{N}\sum\limits_{i=1}^{N}q_{i}-\frac{1}{K}\sum\limits_{i=1}^{K}c_{i} \,\,
$} \\ 
 \scalebox{0.78}{\!\!\!\!$
 q_{i}\!\!=\!\!\sqrt{(\digamma_{i}-\mu_{X})^{T} \big((\Sigma_{X}+\Sigma_{Y})/2\big)^{-1}(\digamma_{i}-\mu_{Y})}
 $}} 
 & $q_i$ represents quality of $i^{th}$ patch; $\digamma_i$ is feature vectors of distorted image; $N$ is length of vector $q_i$; $\mu_{{X}},\mu_{{Y}}$ and $\Sigma_{{X}},\Sigma_{{Y}}$ are mean vectors and covariance matrices of Multivariate Gaussian model of original and distorted images, resp. $c_i$ is object recognition confidence; $K$ objects are there in distorted image.\\
      \hline
    \end{tabular}
    \end{table*} 
\begin{figure}[h]
    \centering
    \includegraphics[width=3.5in]{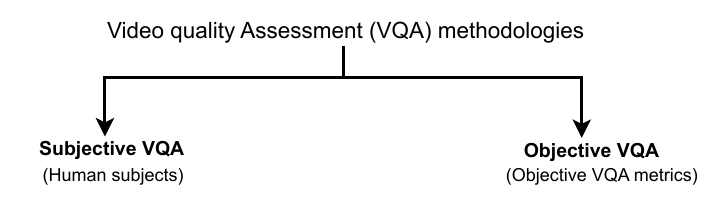}
    \caption{Categorization of VQA methodologies}
    \label{f:mos_dmos}
\end{figure}

The Mean Opinion Score (MOS) or Difference MOS (DMOS) is a well-known measure to evaluate the video QoE. The strategies employed for assessing video quality encompass both subjective and objective approaches, as shown in Fig. \ref{f:mos_dmos}.
A well-defined method for evaluating QoE encompasses the collection of video quality evaluations through the direct involvement of human participants, usually referred to as Subjective Video Quality Assessment.
The method entails undertaking subjective tests by asking human participants about their perspective on the perceived quality. 
Subjective quality evaluation is conducted according to Recommendation of ITU-R BT.500-11 (2000).
Assessors require knowledge of assessment methodologies, grading scales, impairment types etc. Table \ref{tab:opinion} displays a variety of scores and the associated quality. 
Collecting subjective assessments from numerous non-expert assessors is really a difficult process. 
Also, a controlled environment needs to be established to facilitate the optimal viewing of videos. 
 Hence, it is essential to improve the performance of multimedia applications to meet reasonable standards of human perception.
\par 
Alternately, objective Image/Video Quality Assessment (IQA/VQA) metrics can be generated by calculating the artifacts and content features that have an impact on the overall video quality, referred to as Objective Video Quality Assessment. 
Peak-Signal-to-Noise-Ratio (PSNR), Visual Signal-to-Noise-Ratio (VSNR), Singular Value Decomposition (SVD), Structural Similarity Index (SSIM), Multi-Scale SSIM (MS-SSIM),  MOtion-based Video Integrity Evaluator (MOVIE), Spatio-Temporal RR Entropic Differencing (STRRED), Temporal MOVIE (TMOVIE), and Spatial MOVIE (SMOVIE) constitute some of the objective quality metrics~\cite{moqoe}.
These metrics are calculated on the luminance component at each frame with reference to source video in a similar set.
The mean value across all frames yields the objective metrics associated with the video sequence. The objective VQA metrics specifically formulated for 360$\degree$ videos are NCP-PSNR \cite{ncp_psnr}, WCPPPSNR \cite{psnr_vr}, spherical-SSIM \cite{spherical_ssim}, and WS-SSIM \cite{ssim_weighted}. Table \ref{t:acronym} lists the key VQA/IQA metrics, spanning Full Reference (FR), Reduced Reference (RR), and No Reference (NR) types. The VQA metrics are categorized into FR, RR, and NR depending on the information availed from the source videos. FR (e.g., \cite{ssim}, \cite{funque_tr}) necessitates complete information; RR (e.g., \cite{stgreed}), relies on partial information; and NR (e.g., \cite{fovqa, zheng_complete, chipqa, vpniqe}) operates without any dependence on the available source video data.

\subsection{Need for video QoE modeling}
\begin{figure}[h]
    \centering
    \vspace{-2 mm}
    \hspace{-8 mm}
    \includegraphics[width=9.3 cm]{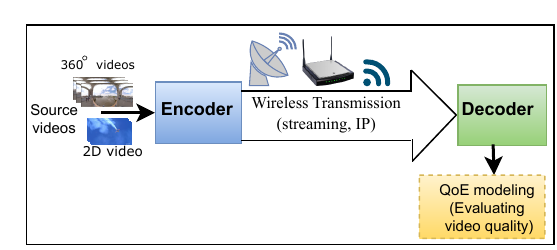}
    \caption{Video transmission over wireless access network}
    \label{f:intro_video_trans}
\end{figure}
As videos pass through multiple processing stages before reaching the end-users, the effect of most of them is to degrade the video quality. Videos can have significant distortions at several stages, such as during processing (image acquisition), compression (encoding), or transmission. At most of the stages, video quality deteriorates. Fig. \ref{f:intro_video_trans} presents the  flow diagram of video transmission process which shows that the videos are subject to several processing stages before and in the course of delivery to viewers that results in potential loss of video quality. 
Recent developments in video coding and compression enable customers to access high-quality video services. The content providers use several compression techniques to offer affordable services. Video encoding causes distortions due to compression.
The application of block-based coding and motion compensation techniques by most video coding and compression standards subjects the decoded videos to one or more compression artifacts. The compression standards, e.g., MPEG 2, MPEG 4, H.263, H.264/Advanced Video Coding (AVC), H.265/ High Efficiency Video Coding (HEVC), and VP9 exhibit blocking, ringing, and blurring artifacts \cite{d2artifact}.
Spatial artifacts that frequently result from encoding comprise false contouring,  mosaic patterns, and contrast distortion. In addition, videos often experience significant temporal artifacts, primarily resulting from transmission across communication networks. 
Video transmission via wireless or IP networks leads to packet loss or corrupted frames that suffer from temporal artifacts \cite{pea265}, like jitter, additive noise, and motion compensation mismatch, which affect individual frames. 
\par This is even more challenging incase of 360$\degree$ videos. 
There are several issues associated with efficient encoding and processing of 360$^{\circ}$ videos ~\cite{evb,i2mb}. It is necessary to project the 360$^{\circ}$ videos onto a flat surface because most filters and coding tools are based on 2D images. At every stage of the 360$^{\circ}$ video processing pipeline, distortion may be introduced, starting with the acquisition of images. Omnidirectional images and videos are typically stitched from numerous cameras \cite{image_stitch}, which can incorporate a variety of peripheral problems. Typical issues include loss of information,  misaligned edges, temporal coherence, ghosting, camera jittering, dominant foreground objects traveling across views \cite{object_travel}, and differences in exposure that are most pronounced at the poles. Temporal discontinuities can also occur in  the videos, like objects appearing and vanishing. Most cameras fail to capture these, while some of them are subsequently reconstructed in the post-processing stage. 

HEVC allows independently decodable tiling of 360$^{\circ}$ videos with less overhead, allowing adaptive streaming according to the user's Field Of View (FoV).
 \par After acquisition of 360$\degree$ video frame, it must be transformed into a planar representation meant for encoding and storage. 
 The Equi-Rectangular Projection (ERP), is the most prevalent projection for 360$^{\circ}$ video. It divides/cuts the visible sphere into several sets of rectangles all sharing the same solid angle. The ERP is used in studies in \cite{erp_intracode}, \cite{erp_spatially_adaptive}, \cite{erp_intraprediction}.
 This projection is inefficient due to distortion at the poles, since more pixels are used to encode the poles compared to the equator. 
 As viewers mostly pay attention in the vicinity of the equator, the poles tend to fall beyond the FoV. The Cubic Mapping Projection (CMP) \cite{cmp_new} is another type of projection in which a cube is built around the sphere with rays radiating outward from the center. The projection mapping is the result of each ray intersecting with only one point on both solids' surfaces. Compared to the ERP, the CMP results in less geometric distortion at the poles than at the edges/corners of a face. The CMP reproduces a sphere better towards the middle of each face, making this intuitive. In \cite{heacmp}, a hybrid version consisting of equi-rectangular CMP is used to improve the coding efficiency. The projection uses  a mapping function and accomplishes a higher level of uniform sampling, thus minimizing the presence of boundary artifacts.
\begin{table}[!t]
 \begin{center}
 \caption{Correlation between objective metrics and DMOS for LIVE \cite{d12sheikh2005live} dataset.}
   \label{t:corr_dmos}
\begin{tabular}{|l|l|l|l|l|l|l|l|l|}
 \hline
 \scalebox{0.62}{Metrics} & \scalebox{0.62}{PSNR}  & \scalebox{0.62}{SSIM} & \scalebox{0.62}{MS{\text-}SSIM} & \scalebox{0.62}{VSNR} & \scalebox{0.62}{M{\text-}SVD} & \scalebox{0.62}{SMOVIE} & \scalebox{0.62}{TMOVIE} & \scalebox{0.62}{MOVIE} \\
 \hline
 	\scalebox{0.62}{SROCC} & \scalebox{0.62}{0.3527} & \scalebox{0.62}{0.5119} & \scalebox{0.62}{0.6749} & \scalebox{0.62}{0.6022} & \scalebox{0.62}{0.3906} & \scalebox{0.62}{0.7607} & \scalebox{0.62}{0.7462} & \scalebox{0.62}{0.7217}  \\  \hline 
\end{tabular}
\end{center}
 \end{table}
 \par Few other projections include- Polar square projection \cite{polarsquare} (similar to barrel projection, but maps the poles to squares), Octagonal \cite{octagonal} projection (can reduce oversampling areas and assemble points into an octagon that can be rearranged into a rectangle prior to encoding), Rotated Sphere Projection (RSP) \cite{rotatedsphere} (unfolds the sphere under two rotation angles and sutures it in the form of a baseball, improving coding efficiency), and offset projection \cite{offsetproject} (higher number of pixels are used to encode areas near the estimated gaze direction, whereas areas at greater angles from it are compressed more tightly; thus, saving bandwidth). 
 \par On generating the planar representation of the  360$^{\circ}$ video, tiling (dividing into tiles) is done, which is a step in the encoding process that can have a considerable effect on the streaming performance. For effective compression of 360$^{\circ}$ videos, the projection and tiling strategy substantially affect the intensity of geometric distortion. It is possible to independently encode the Quality Emphasized Region (QER) of a video at a higher quality compared to the rest parts \cite{optimalviewport}. With Motion Constrained Tiles (MCT), tiled video segments can be efficiently encoded and decoded independently~\cite{evb,maivs}. 
The work in \cite{i2mb} predicts the FoV tiles and viewport with high accuracy by using Long Short Term Memory (LSTM). 
 It is necessary to develop methods that leverage ensemble learning \cite{epass} to enhance viewport prediction accuracy and assign high resolution to tiles where a user’s viewpoint might appear. Determining the right tile size is necessary for conserving bandwidth. The videos can be encoded depending on the popular viewing areas, as in Macrotile \cite{macrotile}. It is critical to identify such larger viewing areas and adjust the macrotiles as per random head movements. This can maximize the quality and reduce the data to be downloaded, resulting in bandwidth and energy savings.
Still, the user's immersive experience will be impacted by any tiles that might be missed while streaming. Thus, video quality needs to be monitored, assessed, and improved.
\subsection{VQA/ IQA correlation with subjective video quality}
Computing the correlation coefficient between the objective metrics and subjective video quality assessment scores indicates/helps assess the relationship between objective and subjective ratings (MOS/DMOS). The correlation coefficient indicates the dependency of DMOS on a given objective score. Correlation coefficient gives the statistical relationship between PSNR and DMOS, SSIM and DMOS, MSSIM and DMOS, Spatial MOVIE and DMOS etc. The computed correlation coefficient values are depicted in Table \ref{t:corr_dmos} for the LIVE \cite{d12sheikh2005live} dataset.
\begin{table*}[!t]
\scriptsize
  \caption{Summary of existing surveys on video QoE modeling and streaming.}
    \label{tab:old_sur}
    \centering
\begin{tabular}{|p{5 mm}|p{5 mm}|p{35 mm}|p{25 mm}|p{9 mm}|p{13 mm}|p{10 mm}|p{10 mm}|p{10 mm}|}
\hline
S.\,No.& \shortstack{Refer-\\ence}  & Major contributions   & Difference  & \shortstack{QoE \\modeling} & \shortstack{ML based \\QoE models } & \shortstack{Adaptive \\streaming} & \shortstack{ML based\\ streaming} & \shortstack{bit-rate\\ prediction} \\ \hline
1 & \cite{video_zhao} & {Survey on QoE assessment and management of video transmission}& {Detailed discussion of QoE models extended to 360$\degree$ videos}  & Yes& No & Brief & No & No\\ \hline
2 & \cite{audiovisual} & {A complete review of subjective and objective audio-visual quality assessment methods, different factors influencing quality and its degradation, datasets for quality evaluation} & {Elaborate discussions on continuous, time varying models with emphasis on ML based techniques}& Yes& No &No & No& No \\ \hline
3 & \cite{contemporary_hossain} & {Survey on video quality metrics (subjective and objective), QoE modelling and methodologies} & {Extension to ML-based 2D and 360$\degree$ video quality modeling}
&Yes &No &No & No & No \\ \hline 
4 & \cite{bentaleb_survey} & {Overview of video coding standards and challenges in HAS, bit-rate adaptation schemes and state-of-the-art related researches } & {Discussion on the state-of-the-art and challenges in 360 video streaming with focus on user-centric feedbacks and ML techniques }& Yes& No  &Yes & No& Yes\\ \hline
5 & \cite{barman_survey} &{Exhaustive review on subjective and objective video quality evaluation methods, QoE modelling for HAS, their influence factors and associated challenges} & {Extension to 360$\degree$ video streaming with  emphasis on ML-based user- centric techniques} & Yes& Brief & Yes & No & Brief \\ \hline
6 & \cite{koug_multimedia} & {Review of subjective  and objective QoE evaluation methodologies,  certain features of QoE assessment for video  streaming, gaming extended reality  }                          & {Discussion about video (normal and  360$\degree$ videos) bit-rate adaptation techniques } & Yes           & Yes                 & No              & No                 & No                 \\ \hline
7 & \cite{measuring_hewage} & {Survey on continuous time varying QoE  assessment models}& {Detailed discussion on ML-based continuous QoE prediction modeling  and streaming} & Yes& Brief& No&No & No\\ \hline
8 & \cite{ad_video_zhou}& {Briefly reviews the quality assessment models for adaptive video streaming that includes subjective studies and objective models along with their performance analysis}& {Extension to 2D and  360$\degree$ ML-based user-centric video streaming} & Yes & Brief & No & No & No \\ \hline
\end{tabular}
\end{table*}
\begin{figure}[!t]
\subfigure[\label{f:frame_cor}]{
\includegraphics[height=4.5 cm]{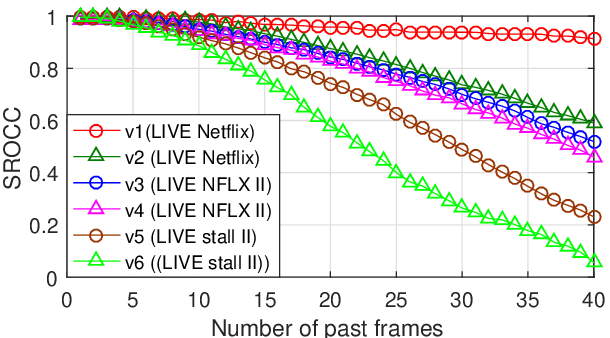} }
\subfigure[\label{f:frame_cor_v}]{
\includegraphics[height=4.2 cm]{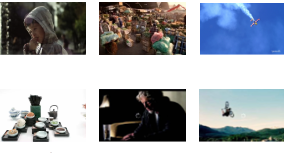} }
\caption{(a) Correlation between QoE of current and past frame indices for the given video samples (b) Video samples (from left to right) (i) v1 (ii) v2 (iii) v3 (iv) v4 (v) v5 (vi) v6}
\end{figure}
 Fig. \ref{f:frame_cor} shows the correlation (SROCC) existing between video quality of frames at current instant and the past instants for videos v1 to v6 shown in Fig. \ref{f:frame_cor_v} from \cite{d59sheikh2005live3, nflx1, mobilestall2}. The correlation gradually decreases for a past frames is further away in the timeline. This correlation is useful for building the continuous, time varying QoE models~\cite{desvq}. 
 \par  The video QoE modeling and user-centric multimedia streaming with special emphasis on ML are less considered in many previous research
works \cite{video_zhao}–\cite{ad_video_zhou}. Table \ref{tab:old_sur} provides a comparative summary of our article with respect to existing surveys in literature. In this survey we holistically cover the ML-based QoE modeling, adaptive and ML-based streaming, as well as bit-rate prediction methods along with the prevalent datasets used in this domain for the conventional and 360$\degree$ video applications.
\section{User-Centric Multimedia Streaming}
\label{s:user_centric}
This section discusses adaptive streaming of multimedia content, wherein the client has complete control over the streaming session and can possibly adapt the multimedia stream to its context, such as network conditions, device capabilities, perceptual quality, etc. The adaptive multimedia streaming  solutions employ an explicit adaptation loop/ logic where clients perform different measurements and push the information towards the server using sophisticated schemes/algorithms.
\subsection{QoE driven multimedia streaming}
In recent times, there has been a notable worldwide increase in the use of multimedia streaming applications. The prevailing major participants in the present global market for these kinds of applications are the Akamai Technologies, Netflix, Apple Inc, Amazon Web Services, and Hulu. 
The rapid evolution of communication networks and widespread usage of smartphones and smart portable devices, with enhanced processing capabilities, facilitate seamless access to multimedia content. 
\begin{figure}[h]
    \centering
    \includegraphics[width=100 mm]{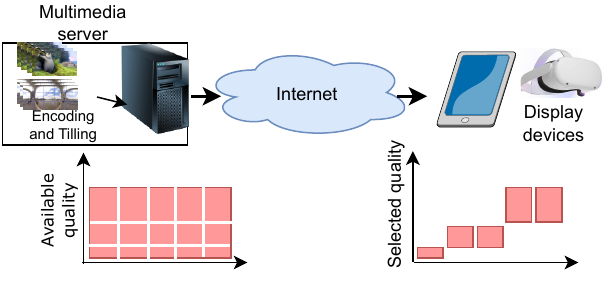}
    \caption{Adaptive video streaming framework}
    \label{f:stream_flow}
\end{figure}
\par DASH is an over-the-top wireless streaming technology that is prominent and effectively enables the adaptability of content delivery in response to changing network conditions. Fig. \ref{f:stream_flow} depicts the visual representation of the streaming concept.
The fundamental concept behind DASH is to break up the multimedia content into small chunks. To have  various interpretations, multiple sets of parameters (i.e., bit-rate, quantization parameter (QP), framerate, resolution) are utilised to encode each chunk. The DASH server hosts the video segments (of certain duration, generally 2, 4, or 10 seconds), where each segment is encoded to have
various representation of quality levels. The obtained representations are logged in the Media Presentation Description
(MPD) file that gives an index for the listed media segments at the server. 
\par DASH effectively manages diverse network conditions through the dynamic adjustment of video parameters. The client-side of DASH is responsible for monitoring various network parameters such as network conditions and buffer size. Based on this information, the client-side determines the appropriate representation of the media chunk to be played next. DASH depends on HTTP and Transmission Control Protocol (TCP) that ensures reliable delivery of data to intended destination. The acknowledgement between the sender and receiver occurs when the receiver notifies the sender on the successful reception of packets, as well as any instances of lost or erroneous packets, prompting the sender to re-transmit them. 
\par Transmission protocols play an important role on  QoE performance in video streaming applications. Both TCP and User Datagram Protocol (UDP) are widely utilized for live video streaming. Nevertheless, numerous live streaming industries are adopting UDP for high-motion videos that typically possess more intricate temporal information. The end-to-end delays associated with UDP are far lower than those of TCP, which is essential for events like live sports.
  UDP doesn't use handshakes, delivery assurances, or duplicate protections to make sure data is correct. Instead, it relies on fundamental mechanisms and checksums to maintain data integrity. The likelihood of certain video distortions, including frame loss and flicker, is increased when UDP is employed. Reliable Multidestination Data Transport Protocol (RMDT) \cite{karpov2023comparative}, an UDP based-transport protocol has proved to be more optimal for certain point-to-multipoint streaming scenarios. RMDT is developed to mitigate common IP network impairments, including packet losses, latency, and jitter, to guarantee reliable transmission. RMDT may serve as a viable alternative to TCP and UDP in future high-bandwidth applications.
  \par The latest version of dash.js \cite{silhavy2022latest}, v 5.0, is a free, open source MPEG-DASH player that functions as a reference client and is useful for academic purposes. The other commercial players in the market include Apple’s AVPlayer, Shaka Player, and HLS.js. The dash.js provides a number of functionalities pertaining to adaptive media streaming. This encompasses the playback of both dynamic and static content via DASH and Smooth Streaming formats. The period transitions operate differently in dash.js, permitting codec alterations, whereas, due to single presentation in Shaka Player, the codecs may remain unchanged. The dash.js v 5.0 supports BOLA \cite{bola}, buffer \cite{huang_bb} and throughput based ABR algorithms.
  \par The codecs govern the multimedia data representation format in  adaptive video streaming applications. There has been an ongoing effort towards developing next-generation coding solutions. The AVC/H.264 is a popular codec that uses a 16x16 macroblock configuration for frame encoding. HEVC/H.265 is a more recent codec that reduces video bitrate by around 50\% relative to AVC/H.264 while maintaining similar subjective quality. The enhanced coding efficiency of HEVC facilitate the 4K video streaming with superior fidelity, high dynamic range (HDR), and wide chromatic gamut (WCG). HEVC/H.265 provides a tiling feature for optimized video streaming, utilizing a 64x64 Coding Tree Unit (CTU) structure to encode each tile. Thus, attaining a superior compression ratio compared to AVC/H.264. The VVC/H.266 is the latest codec that has a coding efficiency of $\thicksim$  50\%  higher for similar subjective video quality compared to HEVC, particularly for HD and UHD video resolutions and $\thicksim$  75\%  higher than AVC \cite{vvc_overview}. The new features in the latest VVC were developed to carry out adaptive streaming with resolution variations, 360$\degree$ immersive video and ultralow-latency streaming.
\par Video quality measurement holds significant importance for video service providers as it pertains to enhancing service delivery and ensuring satisfactory performance under typical network impairments. According to a study conducted by Conviva \cite{conviva}, network operators have experienced significant financial losses as a result of inferior streaming quality. Hence, network providers strive to enhance streaming quality and optimize the whole viewing experience.
The network-related video impairment examination is of the utmost importance due to its detrimental impact on QoE. Different methods for evaluating video quality contribute to the monitoring of quality, thereby ensuring the fulfilment of QoS standards and enhancing the overall performance of the system.
\par The success of a video streaming service or multimedia application is reliant upon the level of contentment experienced by its end-users \cite{d11chen2018cross}. Eventually, it is humans that benefit from the utilisation of these services. The perceptual expectations of end consumers are consistently increasing with the hope of superior quality. QoE depends on many things, such as the analysis of the observer, the type of service, the variety of user tools, and the length of the video. QoE is the end-user's overall impression or appreciation of a service. In \cite{repoprefer}, a survey investigated QoE to be the most desirable user choice when it comes to video delivery, surpassing other factors such as the type of content, ease-of-use,  mobility, timing, and sharing. QoE indicates the extent of a viewer's comprehensive perspective and contentment with various aspects, including the video content, communication networks, service quality, and environmental factors. Therefore, enhancing QoE and satisfying the requirements for superior video quality has emerged as a key goal for content providers, academicians, and video streaming companies. 
\par Due to persistently bad network conditions, video data in the buffer depletes, owing to late arrival of video packets, causing playback to stall, popularly known as rebuffering. The frequent appearance of rebuffering events introduces impairments that have a detrimental effect on the QoE. The frequency and length of stalling, the amplitude and frequency of quality transitions, as well as their temporal positions, are some of the aspects that are taken into consideration for impairment measurements. The estimation of QoE is of utmost importance in order to optimize the delivered quality. Furthermore, compression as well as video bit-rate implemented by DASH, results in degradation of customer's QoE.
\par The adoption of the rate adaptation approach results in a persistent variation in perceptual quality over time, often referred to as time-varying quality \cite{eswaralstm}.
The resultant QoE is influenced by a combination of rate adaptation and rebuffering events. Hence, the video quality encountered by end users throughout a streaming session is characterised by a dynamic variation over time. The continuous assessment of QoE in a streaming session poses a significant challenge due to the presence of non-linear relationships among several elements (for instance, video quality, stalling and frequent bit-rate adaption) that influence QoE. 
In addition to spatial distortions, the quality of streaming video exhibits intricate temporal correlations. Continuous monitoring of QoE throughout a streaming session is crucial for effectively managing shared resources among users and maximizing the perceived video quality.  Moreover, it has the potential to mitigate the loss of quality by effectively adjusting the video bit-rate at the recipient's end.
\subsection{QoE-aware 360$\degree$ video streaming for VR/AR}
\begin{figure}[!t]
\centering
\subfigure[\label{f:hmd_device}]{
\includegraphics[width=32 mm]{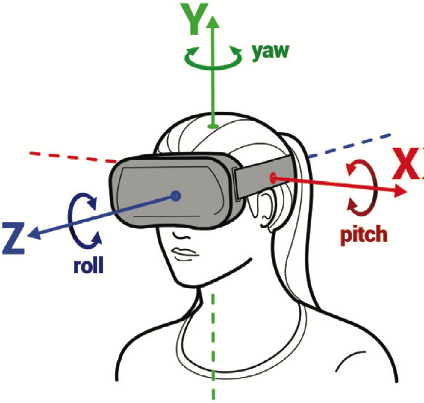} }
\subfigure[\label{f:viewport}]{
\includegraphics[width=32 mm]{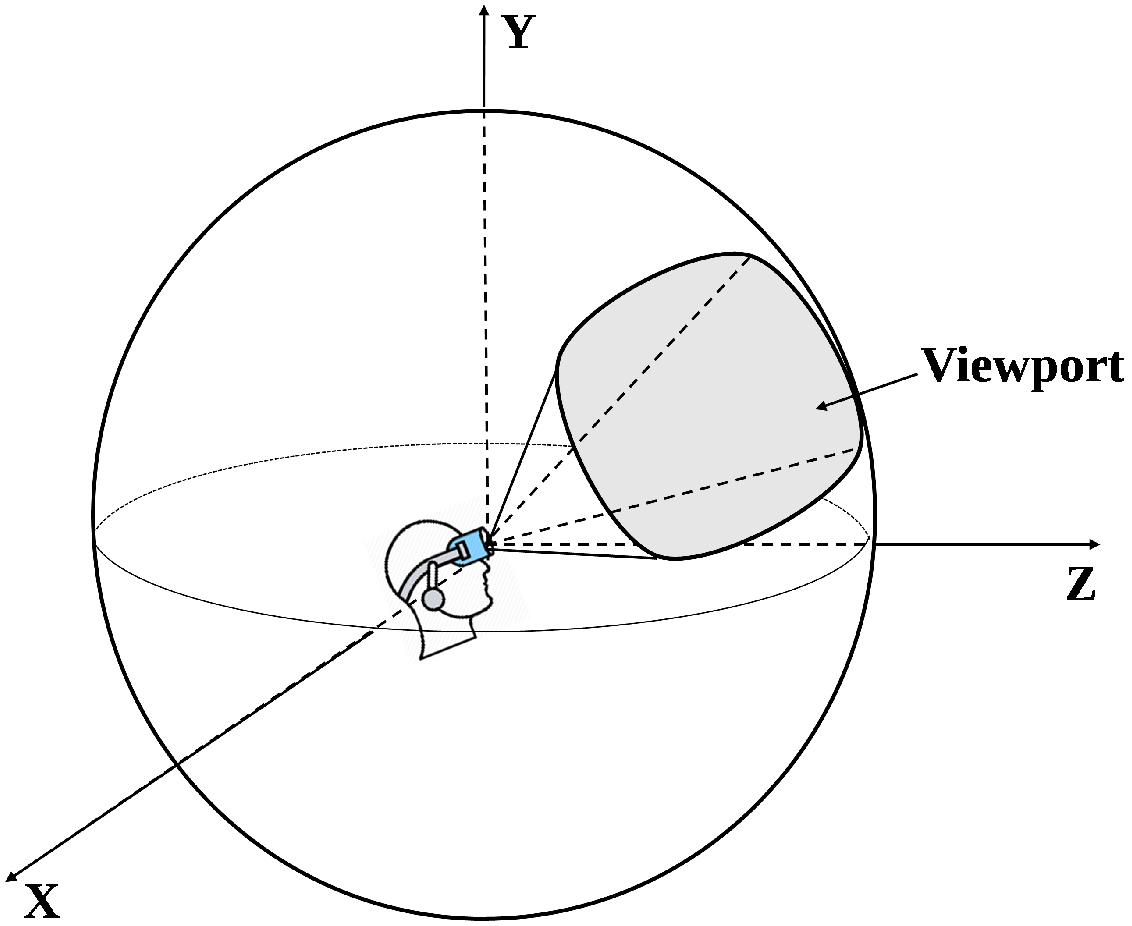} }
\subfigure[\label{f:eqp}]{
\includegraphics[width=30.5 mm, height=24 mm]{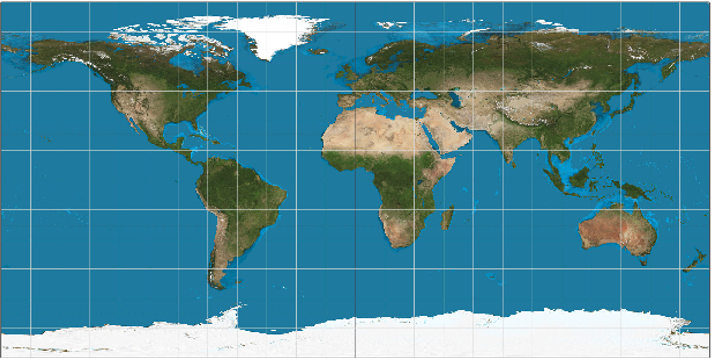} }
\caption{(a) Head navigation direction parameters (b) User’s
viewport (c) Equirectangular projection.}
\label{f:headnavigation}
\end{figure}
The 360$\degree$ video transmission provides viewers with an immersive experience and is a fundamental component of numerous applications, including Metaverse. The 360$^\circ$ video content offers AR/VR application users an immersive experience through the usage of HMD. Users have the ability to rotate their heads in any direction and maintain a seamless, uninterrupted view of the surrounding with these 360$^\circ$ videos. 
 Fig. \ref{f:headnavigation} displays the parameters indicating the direction of head navigation for the users of 360$^\circ$ videos.  Fig. \ref{f:viewport} and \ref{f:eqp} depict the user's viewport and the equirectangular projection of 360$^\circ$  video content, respectively. In equirectangular projection, the sphere is spread out on a flat surface like a cylinder on a 2D sheet.
 \par There is a growing concern over the safety and visual comfort while viewing VR/AR content. A few studies \cite{cybersickness}, \cite{sickness_cog} have reported a range of symptoms including  headaches, trouble concentrating, and dizziness, that attribute to VR sickness, also termed as cybersickness. Contributing factors to cybersickness include optical flow, VR fidelity, user interaction (e.g., navigation methods, controllability), age, gender, VR experience, FoV, latency, and HMD types. Chattha \textit{et al.} \cite{sickness_vr} empirically evaluated motion sickness in VR.
 \begin{figure}[!t]
\centering
\subfigure[\label{f:360_nu}]{
\includegraphics[width=1.9in]{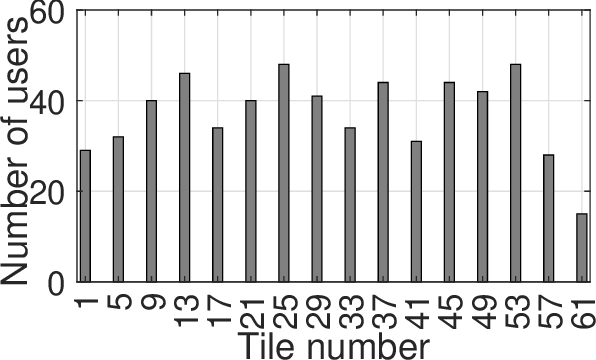} }\hspace{-2mm}
\subfigure[\label{f:360_br}]{
\includegraphics[width=1.9in]{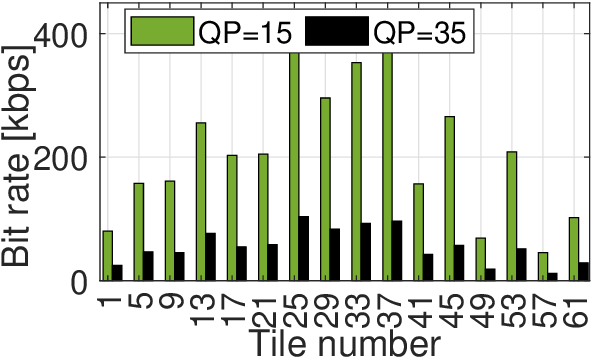} }\hspace{-2mm}
\subfigure[\label{f:360_psnr}]{
\includegraphics[width=1.9in]{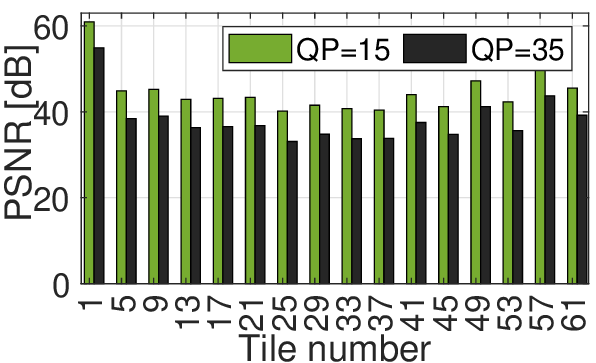} }
\caption{(a) Number of users (b) Bit-rate (c) PSNR of the 360$\degree$ video tiles.}
\label{f:360}
\end{figure}

We study the user head orientations while viewing a 360-degree video using a head mounted streaming device for a set of 55 independent viewings, based on the dataset \cite{hn0,CorbillonDSC:17,i2mb}. Fig.~\ref{f:360_nu} shows the number of users watching the specific tile numbers based on their FoV during the viewing experiment. Certain tiles (57-64) are the least viewed, while others (45-54) are viewed by most of the users. Fig~\ref{f:360_br} and Fig.~\ref{f:360_psnr} show the bit-rate and PSNR (respectively) of individual tiles of the 360$\degree$ video for low (QP=15) and high (QP=35) QP values. A low QP tile has a high bit-rate and a high PSNR than a high QP one. However, these values are not the same for all the tiles of the video at the same QP value, motivating FoV based bit-rate adaptation for efficient 360$\degree$ video streaming.
 
 At higher resolutions, these videos require an exceptionally high bit-rate to deliver an immersive experience~\cite{i2mb}. Excessive bandwidth is wasted while delivering portions out-of-viewport that the end-user never watches. It is difficult to accurately predict the specific content that would interest the viewer in the future playback. This is due to the limited exposure of users to a few 360$\degree$ videos in the past and their fluctuating psychological states and emotional conditions during each playback. Users might adhere to entirely different content depending on moods and mental state.
 Therefore, it is crucial to develop effective strategies for transmitting 360$^\circ$ videos over resource-limited wireless networks while optimizing the viewing experience. Using AR and VR representations, the 360$^\circ$ videos aid in simulating an immersive experience of the real world. 
 \par
\begin{figure}[!t]
	\centering
	\includegraphics[width=4 in]{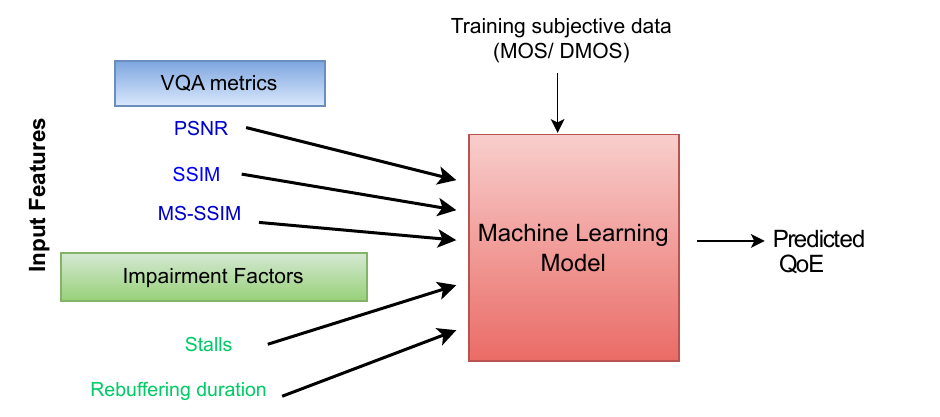}
\caption{{Outline of a ML based QoE predictor}}
\label{f:gen_ml}
\end{figure}

 The most important challenge for successful multimedia streaming is in ensuring the perceptual satisfaction/ pleasure of the end-users.
 Multiple users may demand different immersive experiences since user equipments' ability for FoV prediction might vary depending on various FoV prediction methods. This necessitates the development of novel approaches for synchronous and asynchronous frame structures comprising a basic tier and an enhancement tier for video transmission in multiuser cellular networks, wherein the basic tier is utilized to boost resilience against FoV prediction failures \cite{two_tier}. The users tend to quit the streaming session if the viewing experience falls below a certain threshold level. Accurate QoE estimation/ prediction can help in adapting the content transmission to enhance the viewing experience of the end users. So, a lot of effort is directed towards accessing the 360$\degree$ video quality and improving the immersive experience. The overall video quality is assessed for different segments of the video with short durations ranging from 5-10 seconds. The assessment of video streaming QoE in a continuous manner (on a per-frame basis) is essential for the purpose of regulating the degradation in video quality throughout the entire streaming session. The use of optimized ML models helps in  accurately evaluating the impact of several impairments on the viewers perceptual QoE using different input features (VQA objective metrics and impairment factors). Fig. \ref{f:gen_ml} shows the general design of a ML-based QoE predictor.
\subsection{Intelligent and adaptive multimedia streaming}
\par The DASH server delivers videos at different bit-rates depending on user needs \cite{d5effHAS}. The display devices are capable of adapting various levels of quality  based on the bandwidth that is available. Consequently, this can result in the occurrence of compression or scaling artifacts. For instance, video content that has been encoded at lower resolution may be upscaled to a significantly higher resolution on the viewing device. If the video bit-rate is lower than the available bandwidth, it not only leads to smooth playing but also results in the inefficient use of resources that could otherwise be allocated towards improving the video quality.
\par The standardization of DASH does not mention the execution of the adaptation strategy. For example, how a client can adaptively select the video quality according to the present network statistics and other factors. In order to switch between multiple streams, a controlling mechanism can be configured either at the server/ client side that dynamically predicts and subsequently requests an optimal video segment representation. This request is made considering several factors like network conditions, buffer occupancy, requester's device features etc.
\par Recent studies \cite{bola, sac_abr, abraider, epass, macrotile} report several controlling mechanism, also called Adaptive Bit-Rate (ABR) schemes/
algorithms for DASH. The ABR schemes primarily aim to maximize end user’s QoE by adjusting the video bit-rate/ tiles to fluctuating network conditions and other factors. Choosing the appropriate bit-rate might be difficult due to (i) fluctuations in network throughput (ii) varying video QoE requirements, such as minimizing rebuffers, reducing startup delays, or achieving high bit-rate (iii) challenge of maintaining consistent playback state (play or pause) for multiple clients located in different geographical positions while delivering video content simultaneously (iv) The consequences of bit-rate decisions, such as selecting a high bit-rate that may drain the buffer to a critical level and cause rebuffering (v) confined availability of bit-rate options, which creates a conflict between prioritizing higher video quality and the chance of rebuffering.
\par The ABR algorithms do not properly represent the client’s instantaneous visual perception of quality, nor do they quantify the measure of QoE objectively. Strategies like reducing the number of rebuffering and bit-rate switches may appear logical as they help reduce viewer frustration and ultimately enhance user involvement in the video session. However, they fail to comprehensively record or ensure the optimal (immediate or overall) visual  QoE. Most of the adaptation schemes depend on fixed rules that govern the bit-rate decisions, which needs a lot of parameter tuning and may not generalize well to large network scenarios. The efficacy of Reinforcement Learning (RL) techniques \cite{rl_multiagent, rl_data, rl_linear} demonstrates their viability as a solution in scenarios where prior assumptions about the operational environment are not required.

 \par A 360$^\circ$ video is a bounding sphere that includes all of the surrounding content. A higher bit-rate is needed for smooth streaming at high resolution ($\geq$4K) and high frame rate ($\geq$60 fps) to provide immersive content. The amount of bandwidth needed is many times greater than normal videos of same quality.

It is possible to greatly lower the bit-rate needed for 360$^\circ$ video only by ABR streaming the parts of video which lie in the user’s FoV \cite{nova}. 

This includes breaking down the 360$^\circ$ video into smaller tiled video segments, streaming the segments that lie in FoV at higher resolution/ quality and the remaining at lower quality.
\par Such VR content, even when encapsulated with the most recent codecs, necessitates a significantly higher bit-rate for transmission at UHD or higher resolutions. Despite the growth in network capacity, it is still not adequate to support streaming higher content and fulfill the growing demands of higher video quality \cite{two_tier}. Given the widespread accessibility of display devices that incorporate cutting-edge technological improvements and the significant network bandwidth requirements of streaming users, the primary obstacle in content delivery lies in the development of network-aware solutions aimed at enhancing the overall viewing experience. Multiple service providers might employ varying encoding choices for the same bit-rate~\cite{SinghalIWCMC,SinghalTVT2019}. However, there may be fluctuations in network conditions as a result of factors such as congestion caused by heavy usage during peak hours, the mobility of users, and the extent of network coverage.
\begin{figure}[!t]
	\centering
 \hspace{-6 mm}
	\includegraphics[width=92 mm]{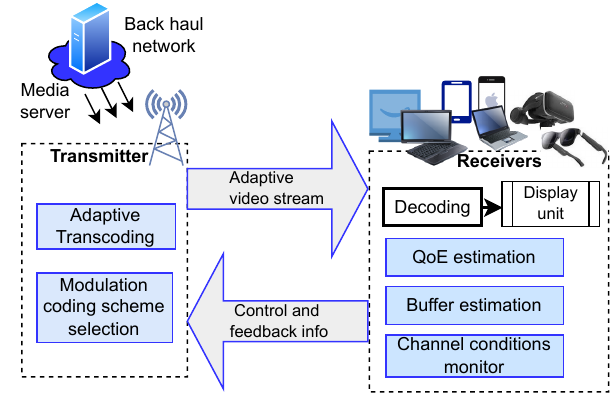}
\caption{{Generalized framework for user-centric multimedia streaming}}
\label{f:general_streaming}
\end{figure}
\par Streaming multimedia content is challenging due to unreliable network conditions and the heterogeneity of end-user devices. Ensuring optimal performance is imperative for streaming applications, since it is crucial to maintain a seamless playback experience with minimal buffering and frequent fluctuations in quality. Devising intelligent schemes can select proper multimedia content based on user-centric feedback. Adaptive schemes/ algorithms help to select appropriate bit-streams based on several user-centric feedbacks, such as buffer occupancy, playback rate, instantaneous throughput of the network, and experienced video quality. The use of Deep Reinforcement Learning (DRL) in bit-rate adaptation schemes facilitates the selection of appropriate bit-rates. 
For streaming the 360$^\circ$ videos, that are of higher resolution, a substantially higher bit-rate is required, even when employing the most recent codecs for encoding. Devising ML based intelligent and adaptive strategies can help reduce the data rate requirement for streaming higher resolution videos. Fig. \ref{f:general_streaming} shows a generalized framework for user-centric multimedia streaming.
\section{Evaluation Metrics and Key Considerations} 
\label{s:evaluation}
Here, we discuss the evaluation metrics and key considerations for the ML-based user-centric multimedia streaming.
\subsection{Objective function and $k$ fold cross validation}
For video quality estimation, the objective/ loss function is MSE
distance between the predicted $(\hat{\textbf{y}}^{'})$ and subjective (ground truth) $(\textbf{y}^{'})$ video quality scores, given as
 \begin{equation}
     \mathcal{L}={\Vert \hat{\textbf{y}}^{'}-\textbf{y}^{'} \Vert}_{2}
 \end{equation}
 When training the QoE framework in a classification problem, cross-entropy is the commonly used loss function, given as 
 \begin{equation}
    \mathcal{L}_{c}=  -\sum_{i=1}^{n}y_{i}^{'} \hspace{2 mm}log \hspace{1 mm} {\hat{y}_{i}}^{'}
 \end{equation}
where, $y_{i}^{'}$ is the $i^{th}$ ground truth,
${\hat{y}_{i}}^{'}$ is the $i^{th}$  predicted quality and $n$
represents the batch size.
To reduce the bias caused by random sampling of training and holdout data samples when comparing the predicted accuracy of multiple ML-based QoE models, studies commonly employ $k$-fold cross-validation. In cross-validation approach, the complete dataset $D$ is initially shuffled. The dataset $D$ is randomly divided into $k$ mutually exclusive subsets $D_1, D_2,\ldots D_k$. Among a collection of k subsets, a combination of $(k-1)$ subsets is chosen for training the regression model, while the remaining subset is chosen as the test set. Therefore, a specific fold is created, comprising a training set and a test set. In this way, $k$ folds can be generated, denoted as $F_{l}$, $\forall l=1,2,\ldots k$. During the process of creating folds, it should be ensured  that each specific subset is chosen only once as the test set.
The cross-validation estimate of the overall performance criteria is determined by taking the average of the $k$ individual performance metrics given as,
\begin{equation}
   \mathcal{T}=\frac{1}{k} \sum_{i=1}^{k}\mathcal{P}_{i}
\end{equation}
where, $\mathcal{P}$ represents the performance measure for each fold.
\subsection{Performance Metrics} \label{s:per_metrics}
This subsection discusses the popularly used performance metrics for assessing the performance of various QoE models.\\
i)\hspace{0.1cm}\textit{Root Mean Square Error (RMSE):} The computation involves taking the square root of the average of all squared errors, which are calculated by comparing each element of the predicted QoE and the actual video QoE of $n$ test videos. \scalebox{0.8}{$RMSE = \sqrt{\frac{1}{n}\sum_{i=1}^{n}(y_{i}^{'}-\hat{y}^{'}_i)^2}$}
where, \scalebox{0.8}{$y^{'}_{i}$} represents the actual score of \scalebox{0.8}{$i^{th}$} test video and \scalebox{0.8}{$\hat{y}^{'}_i$} represents the predicted score of that video.\\
ii) \textit{Mean Relative Absolute Accuracy (MRAA):}
The Mean Relative Absolute Error (MRAE) is calculated by taking the average of all relative absolute errors.

\scalebox{0.8}{$ MRAE = \frac{1}{n}\sum_{i=1}^{n}\left\lvert\frac{y_{i}^{'}-\hat{y}^{'}_i}{y_{i}^{'}}\right\rvert$}.
MRAA is calculated as, 
\scalebox{0.8}{$MRAA = (1-MRAE)\times100 \%$}. \\
iii) \textit{Root Mean Relative Square Accuracy (RMRSA):}
Root Mean Relative Square Error (RMRSE) is calculated by taking the square root value of the average of all relative squared errors between each element of predicted and actual video QoE score. 
\scalebox{0.8}{$RMRSE = \sqrt{\frac{1}{n}\sum_{i=1}^{n}{\left(\frac{y^{'}_i-\hat{y}^{'}_i}{y^{'}_i}\right )}^2}$}. Thus, RMRSA is calculated as \scalebox{0.8}{$RMRSA\!=\!(1-RMRSE)\times100 \%$}.\\
iv) \textit{Spearman Rank Order Correlation Coefficient (SROCC):}
It is a statistical index that gives a measure of monotonicity.
\scalebox{0.8}{$SROCC=\dfrac{\sum_{i=1}^{t} (r_{i}-\bar{y}_{r})(s_{i}-\bar{y}_{s})}{\sqrt{\sum_{i=1}^{t}{(r_{i}-\bar{y}_{r})^{2}}} \sqrt{\sum_{i=1}^{t}{(s_{i}-\bar{y}_{s})^{2}}}}$}
where, $r_{i}$ and $s_{i}$ is the rank of \scalebox{0.8}{$y^{'}_i$} and \scalebox{0.8}{$\hat{y}^{'}_i$} in actual and predicted 
score vector, respectively. \scalebox{0.8}{$\bar{y}_{r}$} and \scalebox{0.8}{$ \bar{y}_{s}$} represents mean of \scalebox{0.8}{$r_{i}$} and \scalebox{0.8}{$s_{i}$}, respectively.\\
v) \textit{Pearson Linear Correlation Coefficient (PLCC):}
PLCC gives an accuracy measurement. It is calculated as follows,
{\small
\scalebox{0.8}{$PLCC=\dfrac{\sum_{i=1}^{n} (y^{'}_i-\bar{y})(\hat{y}^{'}_i-\bar{y}^{'})}{\sqrt{\sum_{i=1}^{n}{(y^{'}_i-\bar{y})^{2}}} \sqrt{\sum_{i=1}^{n}{(\hat{y}^{'}_i-\bar{y}^{'})^{2}}}}$}}
 where, $ \bar{y} $  and $ \bar{y}^{'} $ are mean of actual and predicted score, respectively of test videos. \\
 vi) \textit{Perceptually Weighted Rank Correlation (PWRC):}
PWRC \cite{pwrc} measures the rank accuracy of VQA/IQA metrics, the impact of perceptual significance modifications, and the level of subjective opinion ambiguity by developing non-uniform weighting and adaptive activation methods. The rank accuracy is evaluated using a rank correlation measure that takes into account the confidence level. This measure calculates the area under the curve by collecting PWRC (Partial Weighted Rank Correlation) values within a certain threshold range  
 \scalebox{0.8} {$[T_{min}, T_{max}]$}, given as: {\scalebox{0.9} {$\int_{T_{min}}^{T_{max}} S
(\textbf{y}, \hat{\textbf{y}}, T) dT$}}, where, {\scalebox{0.9}{$\textbf{y}$}} and {\scalebox{0.9}{$\hat{\textbf{y}}$}} represent the actual and predicted score, respectively. \scalebox{0.8}{$S(\textbf{y}, \hat{\textbf{y}}, T)$} is the overall sorting accuracy indicator. \scalebox{0.8}{$T$} represents the sensory threshold.\\
 vii) \textit{Outlier Ratio (OR):} It is uded to measure the consistency.  OR is calculated as ratio of number of outlier-points to number of test videos. \scalebox{0.8}{$OR=\dfrac{\text{Total no. of outliers}}{n}$.}
\section{QoE modeling and Evaluation for efficient multimedia streaming }
\label{s:mul_stream}
Significant studies have been undertaken in understanding and evaluating the impact of several impairments on the viewer's perceptual QoE. The QoE predictors can be modeled to forecast two aspects: (i) the overall video QoE for short-duration segments, and (ii) the continuous, time-varying QoE on a per-frame basis during the streaming session.
\subsection{Overall video QoE prediction}
Assessment of video quality can be done using both subjective and objective methodologies.
Recent studies have examined the subjective evaluation of video quality in \cite{nflx1}, \cite{ugc_sub}, \cite{shang_sub}, \cite{saha2023study}, \cite{hdr_sdr}, \cite{ugc_sub_tr}, while \cite{media_quality}examines the significant difficulties associated with this assessment. 
Subjective assessment methods reflect that tremendous effort has been put in designing the databases for VQA and obtaining the subjective scores. These methods are infact time-consuming and cumbersome. Considering the difficulties and challenges connected with subjective assessment, many approaches in \cite{hidro_vqa, funque, stgreed} resorted to objective assessment methods. Although subjective assessment tasks are difficult, yet their goal is to help in the development of automatic objective VQA models.
\par The subjective assessment methods provide valuable insights into optimizing adaptive streaming performance and user experience. They generate valuable data for studying the impact of various streaming dimensions (including buffering events of various durations, start-up delays, and quality transitions) on end-user experience. These data facilitate the assessment of video quality and QoE models, enabling an analysis of their merits and weaknesses. The subjective studies capture the critical elements of practical systems by integrating real network measurements and/or client ABR schemes. The start-up phase poses the greatest challenge for all ABR schemes, as they have not yet established the video buffer. Thus making them susceptible to network fluctuations that can significantly diminish QoE. The subjective data indicates that viewers can perceive these differences during the start-up phase.  These data illustrate the significance of subjective assessment methods, particularly during the start-up phase, for improving streaming performance. Moreover, continuous scores obtained from subjective assessment methods can be used to train QoE predictor models that can guide ABR schemes.
\begin{figure}[!t]
\subfigure[\label{f:features1}]{
\includegraphics[height=40 mm, width=62 mm]{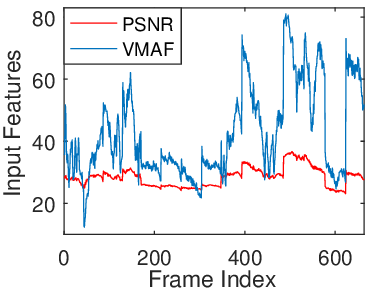} }
\subfigure[\label{f:features2}]{
\includegraphics[height=40 mm, width=62 mm]{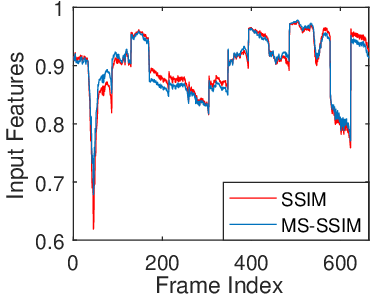} }
\subfigure[\label{f:mos_mask}]{
\includegraphics[height=40 mm, width=62 mm]{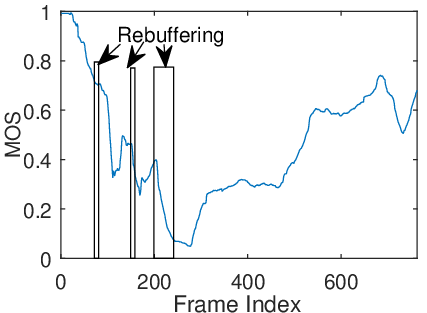} }
\subfigure[\label{f:mask}]{
\includegraphics[height=40 mm, width=60 mm]{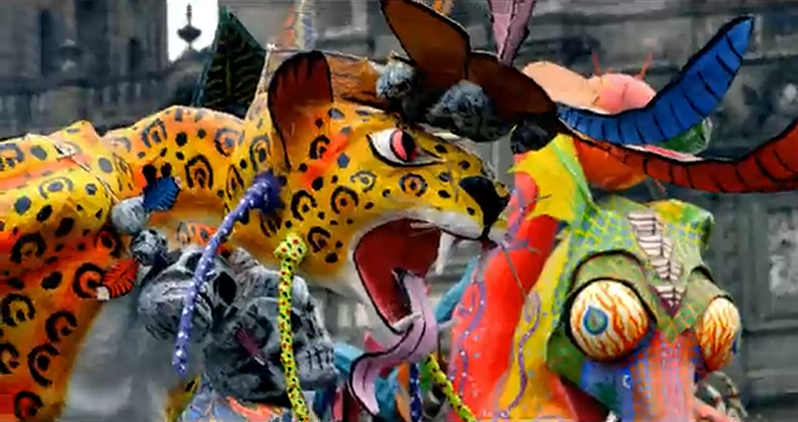} }
\centering
\caption{(a), (b) Variation of different input features vs frame index for the distorted video sample Mask from \cite{d59sheikh2005live3} (c) Variation of subjective MOS vs frame index and indication of re-buffering events (d) Snapshot of the video sample Mask}
\label{f:features}
\end{figure}
\par Objective assessment methods, instead rely on objective VQA metrics, such as, PSNR, Video Multi-method Assessment Fusion (VMAF), FUNQUE \cite{funque_tr}, ChipQA \cite{chipqa}, FOVQA \cite{fovqa}, ST-GREED \cite{stgreed}, VP-NIQE \cite{vpniqe} etc. for predicting the video quality. VQA/IQA gives information of video distortions like  compression, packet loss etc. 
Fig. \ref{f:features1} and \ref{f:features2} shows the variation of different input features (PSNR, VMAF, SSIM, MS-SSIM) vs frame number for the video sample shown in Fig. \ref{f:mask}.  Fig. \ref{f:mos_mask} displays the variation of subjective MOS with respect to the frame index and the occurrence of rebuffering events for the corresponding video sample.
\par With the increasing popularity of VR, there is a growing effort towards developing objective measures that are specifically designed for VR. Adhuran \textit{et al.} \cite{psnr_vr} presented a weighted craster parabolic projection-based PSNR that enhances the existing metric by incorporating rectilinear functions to support VVC for VR-VQA. In \cite{ssim_weighted}, SSIM is exemplified by weighted multiplication in several regions, ensuring that the spherical distortion perceived by the viewer has a linear correspondence with the distortion plane. Croci \textit{et al.} \cite{voronoi} partitioned the 360$\degree$ video into several patches using Voronoi diagram and subsequently applied the 2D methods on these patches to minimize distortion. The traditional objective metrics are not relevant for the VR videos as they do not consider the geometric distortion resulting from the spherical image projection.
\par In \cite{barman_survey}, Barman \textit{et al.}  covers the HTTP Adaptive Streaming (HAS) QoE models, influencing factors that impact QoE modelling, and related challenges.

In \cite{lievens}, Lievens \textit{et al.} presented a MOS predictor in accordance with the user assessment using a parametric equation defined by taking the quality switching, framerate, and rebuffering events into consideration. However, the model's performance is not reported/ verified using subjective assessment. The work in \cite{taha2021automated} presents a methodological framework for assessing subjective QoE considering several factors like video features,  initial latency, segment duration, switching technique, stalls, and QoS indicators. Experimental analysis revealed that objective measurements can be correlated with the most critical subjective parameters of user experience.
Yamagishi \textit{et al.} in \cite{parametric_abs} used separately audio and video quality estimation module that outcomes quality scores per second, which was later integrated in audiovisual integration module into each second audio-visual coding quality. In \cite{tran_novel}, the authors designed a QoE model taking encoded video quality, which is computed per segment by considering average QP, and quality variation into account. 
\par Goring \textit{et al.} \cite{pixelbased} devised hybrid video quality evaluation models that use client-accessible metadata in addition to pixel-based NR to FR models. The hybrid models continue to use the basic information of the distorted videos, like the video codec, resolution, bit-rate, and framerate. Robitza \textit{et al.} \cite{crowdstream} used the ITU-T P.1203 model to measure the streaming quality considering factors such as loading time, stalling, and consumer involvement. For P.1203, Model 0 was utilized, which necessitates the inclusion of codec, bit-rate, framerate, and  resolution data. It has limited applications because of the use of specific codecs and basic inputs. The authors in \cite{directscaling} initially derived the development of the key dimensions of video quality to formulate the entire quality using a new method, known as direct scaling. Their work solely forecasts the overall video quality, rather than focusing on the streaming video quality. The most prevalent models employ various measurement criteria, including  resolution, QP \cite{ewu_tv}, video bit-rate \cite{d3mahapatra2018quality}, and frame rate along with stalling events, to forecast the quality of a video. These models depend on statistical variables such as bit-rate, resolution, QP, and frame rate that are insufficient in capturing the distorted video features and addressing the overall influence on QoE. 
\par 
In \cite{d40tran}, Tran \textit{et al}. examines the impact of QoE deterioration due to factors like starting quality, quality alterations, initial delay, encoding, and rebuffering. Robitza \textit{et al.} \cite{d41robitza} introduced a parametric candidate model, known as P.NATS, for ITU-T P.1203 which adopts a modular approach, combining audio and video quality scores to obtain the final MOS. It obtains coding relevant data per-frame and incorporates changes in quality over time,  as well as stalling. Yet these models do not take into account VQA/IQA measurements as inputs which are perceptually correlated with the quality. The ITU-P.1204.3 \cite{p1204} is the quality assessment  standard for 4K/UHD videos. The developed bit-stream  model \cite{p1204_bit}, based on ITU-P.1204.3 forecasts quality ratings by combining Random Forest regression with parametric model. The RF is used to estimate the residual MOS that the parametric part fails to predict.  The RF uses additional features (average motion/frame, horizontal motion, and frame sizes), apart from the parametric features.
The idea behind this is that the parametric component is not able to adequately encompass the spatial and temporal complexity of the video sequences. It can be noted that in contrast to the conventional FR (\cite{vmaf}) and NR models (\cite{conviqt}) that process decoded frames, which consist of pixel values; the bit-stream based models (\cite{d40tran, d41robitza, p1204_bit}) directly analyze information obtained from the encoded bit-stream. This reduces extra computational load of decoding the videos before evaluating their quality.

\par The work in \cite{hdr_sdr}, examined the effectiveness of various VQA metrics on distortions like scaling and compression while viewing on multiple display devices.  The SpEED \cite{speedqa}  demonstrated superior performance in a particular distortion category, while ST-GREED \cite{stgreed} performed better in another kind of distortion. Various metrics exhibit varying performance for a specific distortion category when viewed on several display devices.  These experimental results revealed that the effectiveness of each VQA metric was constrained in one of the distortion category. This highlights the constraints of individual VQA/IQA measures in considering additional parameters such device display size, video resolution, viewing distance, and other factors that impact video quality.
\par Quality assessment using machine learning techniques has gained a lot of interest in the recent years. Support Vector Machine (SVM) is used in VMAF (0.3.1) \cite{vmaf} to forecast subjective quality based on three elementary input features: motion information, detail loss metric, and VIF.
Further improvements to VMAF has been made in \cite{vmaf_adv} through the use of two SVM models by integrating new features and then merging the results from each of them. The additional features encompass the widely-used VQA metrics (PSNR, SSIM) and are computed independently on the luma and chroma channels, as well as at four distinct resolution scales (2D DWT decomposition). Utilizing such a combination  allows for training on multiple databases. 
In \cite{gaming}, two NR ML methods— Support Vector Regression (SVR) and neural networks— are used to estimate the game video streaming quality.
It is constrained by the use of fundamental feature representations, such as bit-rate, temporal information (TI), and resolution. 
In \cite{mona_book}, Ghosh \textit{et al.} used SVM to estimate the overall QoE. Multi-dimensional QoE estimation for mobile video was achieved using ML models such as SVM, linear regression, bagging tree, and others that were tweaked using the Weka-ML Software tool \cite{d50casas2017improving}. 
Yet, this study was limited by the use of fixed model parameters to achieve the desired performance and was evaluated on a single database.
 The fundamental idea in these works \cite{vmaf_adv}-\cite{d50casas2017improving} is that any basic measure/elementary metric may possess its own advantages and disadvantages in relation to the features of the source content, distortion level, and kind of artifact. The authors integrate basic measures through  ML models, that assigns appropriate weights to each basic metric. This enables the final metric to retain all the advantages of the component metrics and produce a final score that is more accurate.
\par Lekharu \textit{et al.} deployed a Deep Neural Network (DNN)-based model to select the appropriate bit-rate, resulting in total QoE maximization for the user. Tao \textit{et al.} \cite{dnn_datadriven} utilises a Deep Learning (DL) method to estimate the QoE of mobile videos. This is achieved by creating a substantial dataset consisting of more than 80,000 data points, each including 89 network parameters. Nevertheless, gathering all 89 network parameters is a labor-intensive procedure. In DEMI \cite{demi}, Zadtootaghaj \textit{et al.} introduces a Convolutional Neural Network (CNN) (i.e., DenseNet 121) trained using an objective measure that allows it to identify video artifacts. The model is subsequently fine-tuned based on the dataset, taking into account the ratings for blockiness and blurriness. Then, a Random Forest algorithm is applied to  pool frame-level estimations and TI of the videos in order to evaluate the overall video quality. For the evaluation of holographic AR devices, \cite{ar_holo} developed a QoE model that employs a fuzzy inference system (FIS)  to accurately assess the device's performance in a quantitative manner. FIS utilizes fuzzy logic to map the given inputs (i.e., quality of content and hardware, knowledge of environment, user interaction) to an output. The system employs membership functions to precisely specify the degree of fuzziness in a fuzzy collection and calculates output values by applying fuzzy rules. 
In DeepQoE \cite{deep_QoE}, Zhang \textit{et al.} used  a comprehensive framework that utilizes DL techniques (namely, word embedding) to extract generalized features, merge them, and feed them into a neural network for representation learning, with the aim of predicting video quality. DeepQoE demonstrates superior performance when applied to a massive dataset exclusively focused on compression artifacts.  The drawback of DL techniques is that they require a large amount of training data, typically in the scale of thousands or millions. However, the subjective QoE scores collected in complex environments have significantly lower dimensionality and limited public accessibility.
\par In case of user-generated video content uploads, where the reference video is unavailable, different Blind VQA (BVQA)  models have been developed. Compared to conventional feature-based blind VQA  models, DL-based BVQA models have attracted interest recently.
The Patch-VQ \cite{patchvq} extracts the spatial and temporal features using PaQ-2-PiQ and ResNet 3D network, then spatio-temporal pooling is applied in region-of-interest to model the relationship between local-to-global space-time quality.  Ultimately, PatchVQ combines quality scores in a spatio-temporal manner to generate the final predictions. In \cite{li2022blindly}, the authors used transfer learning for BVQA, combining  pre-training on both real IQA databases (to acquire frame-level feature extraction) and large-scale action recognition sets (to acquire motion perception of videos). However, they did not consider the viewing conditions of videos in different environments, which can impact the predicted quality. In \cite{guan}, Guan\textit{ et al.} developed a visual attention module that acquired perceptual quality scores at the frame level. Primarily, the proposed method integrates attention information with spatial-temporal content aspects in a cyclic manner. Next, utilizing a memory attention module which prioritizes quality, the video-level attention-guided characteristics are extracted by changing the dimensions and attention of frame-level depiction.
To generate the best prediction results,  studies in \cite{ugc_sub_tr, rapique} have gained the benefit of using both shallow (statistical) and DL features. It was observed that the DL-based blind VQA models performed significantly better than feature-based models in case of large datasets. 
\par The papers \cite{viewport, fvqa, atlas, block} used techniques of fusing multiple features to combine several objective quality measures.  Azevedo \textit{et al.} \cite{viewport} used various objective measures taken from viewports to forecast the quality of omnidirectional videos.
 The video samples in \cite{fvqa} are first categorized according to their content, taking into account spatial and temporal information values. Multiple FR VQA measures are combined within the groups and given as input to the  SVM to estimate the video quality.  Their research focused exclusively on compression distortions found in the MCL-V dataset, with input limitations restricted to FR. 
The ATLAS \cite{atlas} framework utilizes SVM to forecast video quality based on a limited set of buffering, QoE, and memory-related features. The fusion methods in \cite{viewport, fvqa, atlas} lack any clear objective and are instead merged in a random manner. In addition, metrics that have nearly identical characteristics, such as PSNRhvs and PSNRhvs-M, are not excluded, even if they have only minor differences in performance. This leads to an increase in the total number of inputs and unnecessarily complicates the model. 
\begin{figure*}[!t]
\subfigure[\label{f:dmos_live}]{
\includegraphics[height=42mm,width=46mm]{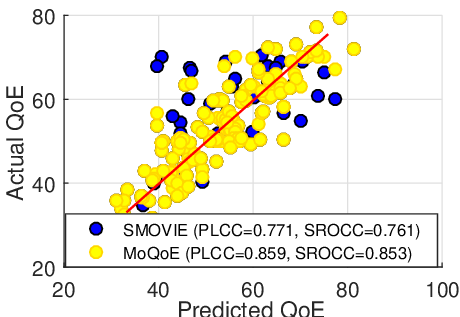} }
\subfigure[\label{f:dmos_waterloo}]{
\includegraphics[height=42mm,width=46mm]{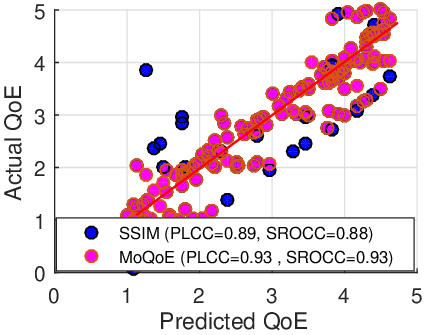} }
\subfigure[\label{f:dmos_waterloo}]{
\includegraphics[height=42mm,width=46mm]{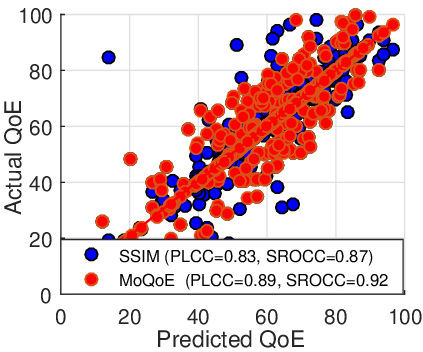} }
\centering
\caption{Scatter plot depicting the correlation between actual and predicted video quality scores for the most effective MO-QoE model for different test ((a) LIVE \cite{d12sheikh2005live} (b) VQEG HD3 \cite{vqeghd3} (c) Waterloo \cite{d17wdb}) datasets. Source\cite{moqoe}}
\label{f:scatter_moqoe}
\end{figure*}
\par
\begin{figure}[!t]
\subfigure[\label{f:f1}]{
\includegraphics[height=40 mm, width=62 mm]{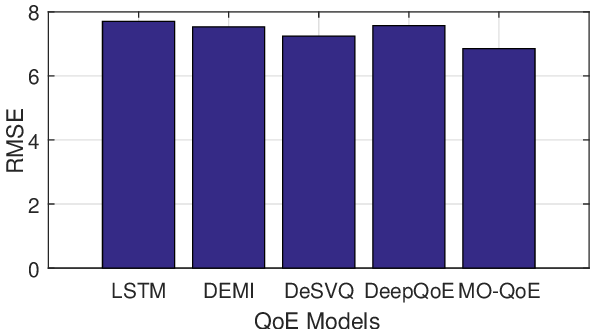} }
\subfigure[\label{f:f2}]{
\includegraphics[height=40 mm, width=62 mm]{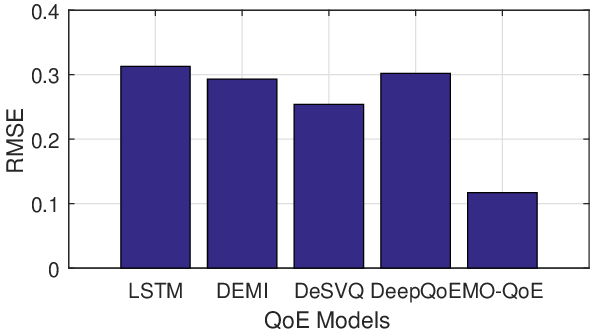} }
\centering
\caption{RMSE of different QoE models (LSTM \cite{eswaralstm}, DEMI \cite{demi}, DeSVQ \cite{desvq}, DeepQoE \cite{deep_QoE}, MO-QoE \cite{moqoe}) when tested on (a) LFOVIA  (b) LIVE Netflix dataset}
\label{f:rmse_models}
\end{figure}

 In MO-QOE \cite{moqoe}, the authors developed a framework that predicts the video QoE using Multi-Feature Fusion based Optimized Learning Models (OLM). The OLMs are the optimized neural network models designed using different combinations of dataset pre-processing techniques, optimization algorithms (Adam/ Batch gradient) and neural network topologies (ANN/FNN). The MO-QOE framework uses a feedback mechanism to determine which features are to be selected in the fusion process using the MFF Algorithm. Adaptive moment estimation and Batch gradient descent techniques are employed to iteratively update the weight and bias parameters of the learning models. The developed MFF algorithm incorporates the feature fusion approach, thus, reducing the model complexity by minimizing the number of inputs. The experimental findings indicate that by employing the MFF-OLM algorithm, adequate performance can be attained with the careful selection of a limited number of features, typically ranging from four to six. Scatter plots of VQA model predictions effectively illustrate model correlations. Fig. \ref{f:scatter_moqoe} displays the MO-QoE \cite{moqoe} model's predictions using five-fold cross validation on different test datasets. It is shown to have a more compact distribution, which aligns with its superior correlation with MOS. The RMSE values obtained for different QoE models when trained on LIVE NFLX II+LIVE Netflix and tested on LFOVIA are shown in Fig. \ref{f:f1}. The RMSE values when trained on LIVE NFLX II and tested on the LIVE Netflix dataset are given in Fig. \ref{f:f2}. The MO-QoE framework performs the best with both the test datasets, obtaining the least RMSE values of 6.853 and 0.117, attributed to its use of optimized models to predict the video QoE.
\par The work in \cite{neurofuzzy} introduced a hybridized  framework consisting of neural network and fuzzy logic. Initially, the neuro-fuzzy framework extracts frame-level features from artifacts and video content, then pools these features using a multi-stage temporal pooling approach. After pooling, these features are applied to a neuro-fuzzy model to predict the quality. Next, the encoding dimensions (e.g., spatial resolution, frame rate, and QP) are adapted, and the acquired features, along with the frame-level features, are applied to the CoActive Neuro-Fuzzy Inference System (CANFIS), followed by Adaptive Neuro-Fuzzy Inference System (ANFIS).
The work in \cite{block} discusses integration of objective metrics at the block level. Their scheme is constrained by a content-driven approach.
The paper \cite{hdr} combines many FR IQA measurements using a greedy strategy that utilises both linear and non-linear regression approaches. The best possible global solution may not necessarily result from such a greedy approach. Additionally, the database under test simply addresses compression artifacts.  
\subsection{Continuous, Time-varying QoE prediction}
 \par  With the advancement of multimedia streaming applications, significant attention is focused on enhancing the seamless and dynamic watching experience.
Multiple studies in \cite{callet_cnn, chen, lfovia, narx, eswaralstm} have conducted research on continuous Time Varying Subjective Quality (TVSQ) evaluation and designing prediction models for continuous time video quality. 
 Crucial elements that affect the quality of watching videos in a streaming scenario are the spatial and temporal characteristics of the reference videos, length and frequency of rebuffering incidents, and the initial delays while starting the video. 
 The studies conducted in \cite{beyondviews,engagement, userquit} examine several factors that influence the user experience during a streaming session. They conclude that when the video quality does not meet the acceptable standards, users are more likely to terminate the session.
 Wu \textit{et al.} calculates the average watching percentage in \cite{beyondviews} utilizing publicly available information, like the video's content, channel, and context, without taking user feedback. 
Lebreton \textit{et al.} forecasts the proportion of users viewing a certain video with reference to time in \cite{userquit}. They examined and found the causes of user disengagement, which include stalls, poor coding quality, and a number of quality-related issues.
\par Callet \textit{et al}. in \cite{callet_cnn} extract several features from MPEG-2 video frames and applied them to CNN to evaluate the continuous video quality. However, their evaluation was limited only to distortions generated by coding artifacts, while distortions due to transmission losses were not examined. 
In \cite{chen}, Chao \textit{et al.} made use of the Hammerstein Weiner (HW) model to get the spatio-temporal features from videos by finding out the Short Time Subjective Quality (STSQ) of video segments.
  STSQ gives an estimate of the perceptual quality of viewers calculated by using VQA metric for every segment of the video. The calculated STSQ were put into a dynamic model which accounts for the hysteresis effects of human conduct and forecasts the TVSQ. Their work is limited in that the TVSQ is only assessed in the context of rate adaptation, without considering rebuffering effects.
  In \cite{bampisrnn}, the authors proposed a kind of Recurrent Neural Network (RNN) using VQA metrics, forecasted video quality ratings, re-buffering, and memory relevant inputs. These inputs are obtained from video at regular intervals and given as input to a nonlinear single hidden layer neural network. 
\par The study in \cite{lfovia} conducts a subjective assessment of streaming videos, as well as QoE prediction using SVR during playback with exponential modeling during rebuffering state. 
In \cite{ghadiyaramlearn}, Ghadiyaram \textit{et al.} determined the time-varying QoE using a multi-stage (where a learner's estimate from one stage is fed into another HW model at the next stage) and multi-learner (where the estimate from each HW model is used to train a different learner) method that examines association between stalling occurrences. The model successfully captures the  interactions between stalling occurrences and subsequently analyzes the spatial and temporal content of the video.
Bampis \textit{et al.} devised a continuous QoE prediction model, known as NARX \cite{narx}, by formulating it as a time series forecasting problem using a non-linear autoregressive model with external outputs. 
NARX's autoregressive memory enables it to consider recent events. Such models are very useful in understanding the human visual system,  which generates the hysteresis effect, in which previous events have a significant influence on the QoE at the present moment. The inclusion of external variables in NARX enables it to represent long-term memory effects, the impact of rebuffering on perceived quality, and current as well as historical video quality. An issue that may occur when employing autoregressive models for real-time QoE prediction is the potential propagation or amplification of prediction errors when the estimated outputs are given back into the prediction engine.
\par Eswara \textit{et al.}, used a network of LSTM \cite{eswaralstm} to record the temporal dependencies for forecasting the continuous, time varying QoE of streaming videos. The LSTM network receives inputs via assessing the STSQ using VQA metrics. Also, it incorporates rebuffering-related inputs at a reasonable time step, which is measured per second. The unique configuration of the gating mechanisms and memory cell in LSTMs allows them to successfully capture the temporal intricate dependencies in QoE modeling. Also, it helps to overcome the limitations of ordinary RNNs, like the issue of vanishing gradients.  The unidirectional LSTM architecture just takes forward dependencies into account. There is a chance of overlooking valuable information. So, the work in \cite{ducbidirectional} employed a Bidirectional LSTM model to process the inputs derived from VQA metrics, memory-relevant data, and rebuffering. The Bidirectional LSTM considers both forward and backward dependencies, thus improving the preduction accuracy.
\par Eswara \textit{et al.} \cite{eswaralstm} exclusively employed the LSTM network, whereas Duc \textit{et al.} \cite{duccnn} solely utilized the Temporal CNN (TCNN), employing a nearly same set of inputs as \cite{bampisrnn}. Duc \textit{et al.} \cite{duccnn} used the TCNN to overcome the computational intricacies of LSTM networks.  Because of the sequential processing feature in its architecture, LSTM has a high computational complexity. This gives rise to a question of how well it would function on devices with low computing power. TCN has the feature of  parallel computing, which offers benefits in terms of modeling and computation. Also,  TCN utilizes dilated causal convolutions to effectively capture the temporal dependencies. Due to the sequential nature of LSTM's recurrent structure, the model was unable to properly harness the parallel computing capabilities, resulting in an increase in computational cost. In \cite{ulsupervised}, the authors examined DASH video application in a practical emulation environment that relies on actual 5G traces (in both static and mobility conditions) to evaluate the QoE of three ABR algorithms (i.e., hybrid, buffer, and rate-based). Then they proposed a supervised ML classifier to forecast user contentment by taking into account network characteristics, including Round Trip Time (RTT), throughput, and the quantity of packets.
\par In M-3R \cite{m3r}, the framework is designed using LSTM networks, which are able to accurately simulate the temporal dynamics of streaming video quality under 3R settings. The "3R"- settings combine the perceptual contributions of numerous FR, RR, and NR VQA metrics. The 3R input features (per frame basis) are given to the LSTM networks because they can correlate highly with perceptual quality. A frame based approach is utilised in M-3R learning framework, as their analysis revealed that QoE is more sensitive to the correlation existing between the frames. In DeSVQ \cite{desvq}, an integrated framework is designed comprising of CNN and LSTM networks. It uses a two- stage feature processing approach, with the first stage processing high level spatio-temporal features and the second stage processing low-level features examined by the VQA metrics. In stage-I, in order to extract the high level spatio-temporal features from the distorted videos, CNN is quite effective. In stage II, the non-linearities and temporal dependencies associated with QoE changes are captured by the LSTM networks. A linear layer combines the output from both stages and feeds it to the decision trees. This is how the framework maps the video features to continuous quality scores. 
 Yang \textit{et al.} proposed a Light-weight QoE model, LiteDC \cite{litedc} by integrating temporal Dilated Convolution network with a tailored pruning strategy for multi-device video streaming. Dilated convolutions fundamentally include the incorporation of ``holes" or intervals between the elements of the convolutional kernel, hence expanding the kernel's engagement with the input data while preserving a streamlined model architecture.
This allows the network to effectively capture long-range dependencies without a corresponding increase in computational complexity, as in LSTM networks.
\begin{figure}[!t]
\centering
\subfigure[\label{f:stream1}]{
\includegraphics[width=95 mm]{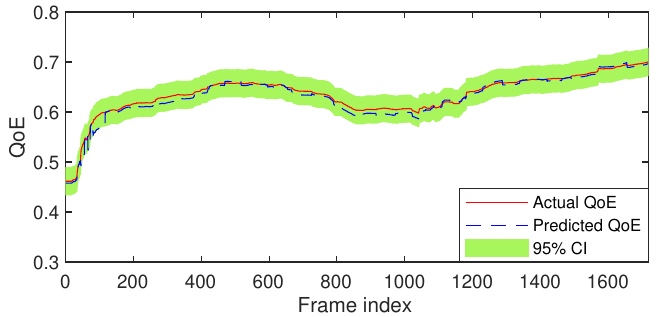} }
\subfigure[\label{f:stream2}]{
\includegraphics[width=50 mm, height=44 mm]{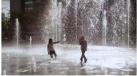} }
\small
\caption{(a) Continuous video QoE prediction result on test video sample from LIVE Netflix \cite{nflx1} shown in (b). CI represents the confidence interval.}
\label{f:stream_nflx1}
\small
\end{figure}
\par  The existence of such ML-based continuous time varying QoE evaluation models assist in taking decisions on the adaptation of video streams. During adaptive HTTP streaming, a decision is made regarding which chunk should be selected for the next delivery. Such kind of modeling incorporated with hysteresis effect and other non-linearities, aid in selecting the segment with the best quality.  Fig. \ref{f:stream1} displays the outcome of the continuous video quality prediction model DeSVQ \cite{desvq} on a test sample \ref{f:stream2} taken from the LIVE Netflix dataset. It can be seen that the predicted QoE very closely follows the actual QoE.
 \par Table \ref{t:qoe_models} provides a review of ML-based continuous, time-varying as well as over the entire video QoE modeling.  The ($\uparrow$) and ($\downarrow$) indicates the increase and decrease in value respectively.
\begin{table*}[!htb]
\scriptsize
\centering
\caption{Summary of the reviewed ML-based QoE models. Cont.: Continuous}
\label{t:qoe_models}
\begin{tabular}{|p{9 mm}|p{22 mm}|p{32 mm}|p{27 mm}|p{22 mm}|p{15 mm}|p{8 mm}|}
\hline
\!\!\!Category & ML Technique & Influence Factors & Accuracy & Datasets used& Reference paper& \shortstack{Pred-\\iction}\\ \hline
\multirow{15}{*}{\shortstack{QoE \\\hspace{-0.5mm}modeling}} &     SVM         & STRRED, stalling(length, duration, frequency), last bit-rate drop time, impairment duration    &  lowest prediction uncertainty than existing      &   LIVE Netflix\cite{nflx1}, Waterloo\cite{d17wdb}       &  ATLAS \cite{atlas} & Overall  \\ \cline{2-7}
&   SVM, bagging tree, linear regression  &  duration, frequency \& starting time of stalling, playback delay   &  reduces prediction errors between 25\% and 50\% than existing        &    LIVE avaasi Mobile\cite{avaasi}         &   \cite{d50casas2017improving} & Overall  \\ \cline{2-7}
&    CNN (DenseNet 121) + Random Forest       &  VMAF, PSNR, blockiness \& blurr ratings  &   same as existing models (on gaming data)      &  KUGVD\cite{barman2019no}; LIVE Netflix, LIVE NFLX II\cite{d59sheikh2005live3}   &     DEMI  \cite{demi} & Overall  \\ \cline{2-7}
 &    3D CNN      &  video, text, categorical info (resolution), continuous values  & 35.71- 44.82\% (on small data);90.94\% (on large dataset)        &      WHU-MVQoE2016 \cite{zhang2016whu}, LIVE Netflix   &      DeepQoE  \cite{deep_QoE} & Overall  \\ \cline{2-7}
 & Fuzzy-logic   &  content quality, hardware quality, environment understanding and user interaction.  &  3.89-5.791  (RMSE)   &   Young Conker, RoboRaid &   \cite{ar_holo} &  Overall \\ \cline{2-7}
 &   Optimized Learning Models (ANN, FNN)       &   VQA metrics+ impairment factors (multi-feature fusion strategy)  &       $\approx$2 to 3$\times$ than existing    &\!\!\!LIVE\cite{d12sheikh2005live},\,Waterloo, VQEG\,HD3\cite{vqeghd3}, VQEG\,HD4\cite{vqeghd3}, LIVE Netflix, LIVE\,NFLX\,II&    MO-QoE  \cite{moqoe} & Overall  \\ \cline{2-7}
  &  convLSTM, ResNet 50      & video (spatial features, spatial-temporal attention-guided features)  &  0.894 (SROCC), 0.902 (LCC) (on CVD2014); 0.836(SROCC), 0.834 (LCC) (on LIVE- Qualcomm)     & VD2014,
LIVE-Qualcomm, KoNViD-1 k, LIVE-VQC, Youtube-UGC    &      \cite{guan} & Overall  \\ \cline{2-7}
 &  Neuro-fuzzy (ANFIS, CANFIS)      &  Artifacts (blockiness, blurr, noise)+ content characteristics  &  0.94-0.97 (LCC), 0.26-5.2 (RMSE)       &     LIVE\cite{d12sheikh2005live}, LIVE MVQ\cite{live_mvq},  CSIQ\cite{csiqvqa},\,IVP\cite{ivp_data}  &      \cite{neurofuzzy} & Overall  \\ \cline{2-7}
 & Non-linear autoregressive models              & VQA metrics, playback status, time since last video impairments &       Outperforms HW model in RMSE and outage rate   &       LIVE Netflix   &       NARX \cite{narx}  & Cont.  \\ \cline{2-7}
& RNN \& feed-forward multi-layer neural network  &  Estimated video quality, rebuffering, recency   &   performance improvement by 5-10\% in OR      &  LIVE-avaasi Mobile,LIVE Netflix,LIVE HTTP video streaming \cite{chen}       &      Bampis \textit{et al.} \cite{bampisrnn}  & Cont.  \\ \cline{2-7}
 & Linear Regression        &  current short time \&  previous time varying video quality    &    39-41\%$\downarrow$  rmse      &   LIVE HTTP video streaming &  Eswara \textit{et al.} \cite{eswaralinear} &  Cont. \\ \cline{2-7}
& SVM           &  Current \& previous time slot's video quality, rebuffering frequency, \& duration  &    34.1\%$\uparrow$ relative LCC gain than eTVSQ \cite{etvsq}     & LFOVIA\cite{lfovia}, LIVE HTTP video streaming       &       Eswara \textit{et al.} \cite{lfovia} &   Cont. \\ \cline{2-7}
 & LSTM & short time video quality, playback indicator, duration since last rebuffering  &  4-29.1\%$\uparrow $ relative LCC than \cite{lfovia, chen, nlss}       & LFOVIA,LIVE Netfix,Mobile stall II\cite{mobilestall2},LIVE HTTP video streaming    &       Eswara \textit{et al.} \cite{eswaralstm} &   Cont. \\ \cline{2-7}
& Temporal convolutional network &  STRRED, playback indicator, \#rebuffering, time since last impairment   &   1.15-38.7\%  $\uparrow $ relative LCC than \cite{eswaralstm, narx, nlss, shi2019con}    &   LFOVIA, Mobile stall II, LIVE Netflix     &       Duc \textit{et al.}  \cite{duccnn} & Cont.  \\ \cline{2-7}
 &   Bidirectional LSTM        &   short time video quality, playback indicator, duration since last rebuffering, number of rebuffering     &   11.4-43.9\%  $\uparrow $ relative LCC than \cite{eswaralstm, nlss, narx}    &   LIVE Netflix       &       \cite{ducbidirectional} & Cont.  \\ \cline{2-7}
 &    Decision
Tree, Multi-linear and Random
Forest Regression   & RTT, throughput, packet count per video segment    &    87.6\% (static) and
79\% (mobility scenarios)  accuracy by Random Forest     &    \cite{quinlan}, \cite{raca5g}    &   Mustafa \textit{et al. }  \cite{ulsupervised} & Cont.  \\ \cline{2-7}
 & LSTM        &   VQA metrics (FR+RR+NR)  &    2.5-20\% relative PLCC
gain     &      LIVE Netflix, LIVE NFLX II, Mobile stall II  &       M-3R   \cite{m3r} & Cont.  \\ \cline{2-7}
& LSTM+CNN+ XGBoost         &   distorted video frames+ VQA metrics  &        4.8-20\% relative PLCC
gain    &   LIVE Netflix, LIVE NFLX II, Mobile stall II        &       DeSVQ   \cite{desvq} & Cont.   \\ \cline{2-7}
 &  Temporal
dilated convolution network      &  VQA metrics, time since last rebuffering, playback indicator, rebuffering duration from beginning of playback, VQA switch &    6.4\%$\uparrow$ in prediction accuracy  than baseline &  LFOVIA \cite{lfovia}, MCQoE \cite{mcqoe} &   LiteDC   \cite{litedc} & Cont.  \\ \hline
\end{tabular}
\end{table*}
\section{Intelligent and Adaptive Video Streaming Techniques}
\label{s:iavs}
This section provides a review of the broad research efforts directed towards the design of bit-rate adaptation schemes accounting for perceptual optimization of streaming video. 
\subsection{Video streaming schemes/strategies}
\par A survey undertaken by Seufert \textit{et. al} in \cite{seufertsurvey} categorises the streaming video QoE influencing factors into perceptual (i.e., time varying video quality, stalling frequency and its duration, initial delay) and technical (i.e., video content and encoding parameters, segment size and duration, adaptation logic/algorithm, hardware/software employed in video streaming system) that directly or indirectly affect the QoE. Impact of such factors and their techniques of measurement can be found in \cite{juluri_vod}. These factors are challenging when considered for obtaining a trade-off between the conflicting objectives (e.g., maximizing bit-rate vs. minimizing rebuffering).
\par The paucity of concrete quantitative methods for measuring streaming video QoE itself is an extensive area of research. There exists not an exclusive set of QoE measurement metrics that can comprehensively measure QoE. Works in \cite{lfovia}, \cite{hdr_sdr} assess QoE using FR objective metrics that do not capture temporal distortions well \cite{nflx1}. 
RR (e.g.,  Video-RRED) and NR (e.g., VP-NIQUE \cite{vpniqe})  metrics are not effective in evaluating perceptual video quality subject to streaming relevant distortions, like rate adaptations,  rebuffering or mixed interplay of both. These methods are followed by DASH adaptation schemes in \cite{seufertsurvey}, \cite{dash_sets}. In D-DASH \cite{d-dash}, Gadaleta \textit{et al.} selected SSIM (that needs complete info of uncompressed segment) to access the instantaneous video quality which was pre-computed on the server side and added in the MPD. However, this increases the computational load on the server side. In \cite{bampis_towards}, Bampis \textit{et. al} designed a database that has subjective QoE scores collected in adaptive streaming scenario. Here, VMAF was used for quality computations that was saved in a chunk map and presented for bit-rate adaptation on client side. It is not viable to compute the FR and RR metrics on the client side as they require complete/ partial info of the reference segment. This will deviate from the realistic assumption of the absence of reference information on the client side. Although NR metrics can be used, the consequences of inaccurate estimates is a matter of concern. 
 \par Online video streaming sessions need consistent track of end user's QoE under dynamic network conditions; so that the perceived video quality does not fall below an acceptable level. In-depth reviews of the parametric, bit-stream, and hybrid models are analyzed and discussed in \cite{barman_survey}.
 Parametric and bit-stream models exclusively reflect the QoE as a function of influencing elements, such as initial loading delay,  switching amplitude, stalling frequency and duration, QP, and framerate. These models use various mathematical functions, such as linear, exponential delay, curve fitting, and ordinal logistic, to describe the relationship between QoE and these parameters. 
 Taha \textit{et al.} \cite{taha2023smart}      developed a streaming framework based on adaptive quantization. They established a relation between the QP in H.264 and H.265 codecs and the QoS in 5G wireless networks. They simulated packet loss characteristic of the network to assess the influence of QP on the delivered video quality utilizing both objective (PSNR and SSIM) and subjective (DMOS) quality metrics. The framework can automatically estimate the QoE by identifying the packet losses. Thus finding the optimal QP value to improve end-user QoE.
  The DASH framework in PENSIEVE \cite{pensieve}, Model Predictive Control (MPC) \cite{mpc}, and Buffer Occupancy based Lyapunov Algorithm (BOLA) \cite{bola} employ several QoE measures, where QoE is defined as a linear and log function of segment bit-rate and rebuffering time. 
 QoE metrics defined in this manner (i.e., as simple representations of influencing factors) are not capable of addressing the distortions and long-lasting dependencies inherent in time-varying QoE.
 \par A detailed survey of different rate adaptation algorithms for DASH is presented in \cite{bentaleb_survey}.
 Bit-rate adaptation schemes are categorized as client-based, network-aided, server-based,  and hybrid adaptation, according to the specific system entity where the logic is implemented.
 In \cite{multiuser_dash}, Ozfatura \textit{et al.} proposed an optimal network-supported multi-user DASH
video streaming. Client based adaptations chose suitable bit-rate by adapting to one or more parameters such as bandwidth variations, buffer size, etc. 
PiStream \cite{pistream}, an adaptation scheme allows DASH clients in LTE network to determine the accessible bandwidth by virtue of a resource control component. In \cite{dash2m}, Xiao \textit{et al.} devised a DASH to Mobile (DASH2M) streaming strategy using HTTP/2 server push along with stream completion properties for adjusting bandwidth so as to reduce client's battery usage while improving QoE. Huang \textit{et al.} \cite{huang_bb} put forward Buffer Based (BB) rate adaptation algorithm where the client picks bit-rate of the next segment depending on buffer occupancy to avoid excessive rebuffering. Jiang \textit{et al.} \cite{festive} devised a Rate Based (RB) adaptation scheme where the efficiency can be improved in HTTP adaptive video streaming by choosing the maximum possible set of bit-rates in order to optimize the end user experience. BOLA \cite{bola}, an online control logic frames adaptation of bit-rate as a utility maximization problem that includes two main elements of QoE, i.e., video's average bit-rate and rebuffering duration. 
\par In \cite{quetra}, Yadav \textit{et al.} introduced a method called QUETRA (QUEuing Theory approach to DASH Rate Adaptation) which models the client as an M/D/1/K queue. This model allows for the computation of the desired buffer level based on parameters such as bit-rate option, buffer space, and network throughput. Nevertheless, their assessment was restricted to only four network profiles, consisting of two profiles from the DASH Industry Forum \cite{dash_industry} and others from HSDPA \cite{hsdpa_url} dataset. Yin \textit{et al.} devised a MPC \cite{mpc} technique which utilizes a throughput estimator to forecast the anticipated throughput for the subsequent five chunks.
 They developed fastMPC, a method that combines throughput and buffer capacity predictions to generate decisions that approach performance levels near to that of MPC.
Such control theoretic techniques have constraints on development time, which must be significantly shorter due to the need for a complete redesign when altering model parameters.
 This approach is similar to dynamic programming, and the effectiveness of the adaption logic may be compromised if the throughput estimator is not correct, leading to suboptimal decisions. Furthermore, the MPC method does not possess a comprehensive assessment of its performance in real-world tests. 
The bandwidth-reliant adaptations in \cite{pistream, dash2m} often have restrictions in obtaining improved QoE due to the lack of accurate methods for predicting bandwidth, which leads to frequent buffer underflow. Buffer-based adaption methods in \cite{bola,huang_bb} often have instability issues when there are prolonged variations in bandwidth.
 \par A number of interesting studies in \cite{smdp2}, and \cite{mDASH} have utilized the Markov Decision Process (MDP) to develop bit-rate adaptation techniques. These techniques employ dynamic programming to determine the optimal approach for adaptation. The study in \cite{smdp2} used MDP to simulate perceived content-aware bit-rate adjustment for adaptive streaming issues.  Next, they formulated a segmented value iteration rate adaption approach to address this issue, which disaggregates an MPD session into several sub-sessions and applies the value iteration method to determine the ideal solution for each interval.
 In mDASH \cite{mDASH}, Zhou \textit{et al.} introduced a method  that incorporates a bit-rate adaption algorithm based on MDP optimization that takes into account buffer measurements, bit-rate stability, and bandwidth scenarios as state variables.
 They suggested a pseudo-greedy heuristic approach that suffers from computational burden. The complexity of such models is prohibitively great to be solved in real-time. 
 \par In order to address the primary challenges of dynamic programming, namely the computational burden and the need for prior knowledge of network conditions and video content, various studies have employed RL techniques. The RL agent acquires experience through its interactions with the environment and constructs an ideal policy. 
PENSIEVE \cite{pensieve} is a learning system that utilizes data collected by DASH clients, including throughput traces and buffer occupancy, from previous segments to make observations. PENSIEVE employs an actor-critic approach to learn its policy, utilizing the fundamental gradient ascent method.  The paper \cite{ictc2022} utilizes an Online RL (ORL) technique, similar to PENSIEVE, which is based on the fundamental policy gradient method.
 These methods have limitations, specifically, they are not efficient in terms of sample usage as they require collecting new samples for practically every policy update.
Comyco \cite{comyco} utilizes imitation learning instead of previous RL-based approaches to train the neural network. The reason is that the near-optimal policy may be accurately and immediately approximated from the present state in the ABR scenarios. Also, the gathered expert policies can facilitate rapid learning of the neural network. Comyco's objective is to prioritize the selection of bit-rates that offer superior perceptual video quality, instead of high video bit-rates.
For mobile edge computing-supported short video applications, \cite{rbc} uses a DRL strategy to optimize video quality benefit while minimizing bearer costs as well as latency penalties. The method uses a policy gradient approach to pick video quality levels without the need for explicit computation of the actual video quality. Simply linking higher resolution to improved video quality does not provide an accurate representation of the experienced video quality as QoE is affected by a number of factors.
\begin{table*}[]
\small
\centering
   \caption{Overview of comparison between the state-of-the-art ABR schemes}
   \label{t:related_compare}
    \smaller
\begin{tabular}{|l|l|l|l|l|l|}
\hline
Scheme   & \shortstack{Buffer \\ adaptation} & \shortstack{Bandwidth \\ estimation} & Network       & \shortstack{QoE \\ Optimization} & QoE modeling methodology       \\ \hline
BB \cite{huang_bb}      &  \checkmark             & \ding{53}        & Fixed network & \ding{53}            & -                             \\ \hline
RB \cite{festive}       & \ding{53}            &  \checkmark          & Fixed network & \ding{53}            & -                             \\ \hline
MPC \cite{mpc}      &  \checkmark             &  \checkmark          & Simulated            &  \checkmark            & Linear equation               \\ \hline
PENSIEVE \cite{pensieve}  & \ding{53}            & \ding{53}         & Simulated, WiFi          &  \checkmark             & Linear, Log equation          \\ \hline
BOLA \cite{bola}     &  \checkmark             & \ding{53}         & Simulated             &  \checkmark             & Average playback interruption \\ \hline
D-DASH \cite{d-dash}  &  \checkmark             & \ding{53}         & Simulated           &  \checkmark            & Objective metric (SSIM)             \\ \hline
ORL \cite{ictc2022}    & \ding{53}             & \ding{53}         & Simulated            &  \checkmark            & Linear equation               \\ \hline
SAC-ABR \cite{sac_abr}   & \ding{53}            & \ding{53}         & Simulated            &  \checkmark            & Linear, Log equation               \\ \hline
ABRaider \cite{abraider}   &  \checkmark             &  \checkmark        & \shortstack{Simulated, Experi-\\ mental (WiFi,4G)}          &  \checkmark             & Linear, Log equation               \\ \hline

\end{tabular}
\end{table*}
\par In SAC-ABR \cite{sac_abr}, Naresh \textit{et al.} proposed an off-policy method for ABR streaming which utilizes Soft Actor-Critic (SAC) driven DRL. The objective of SAC-ABR is to optimize the entropy while simultaneously improving the expected benefits, leading to a more balanced approach between exploration and exploitation.
Guo \textit{et al.} introduced a DRL approach in their paper \cite{guo_buffer} to address the issue of streaming in small-scale wireless networks by incorporating a buffer-aware strategy. 
To determine the productive video streaming period when neither an underflow nor an overflow of playback occurs, they created a reward function. 
 Their proposal involves framing the issue of allocating bandwidth and managing buffers (using MDP) simultaneously in order to optimize the duration for which each user can efficiently stream videos. Simply mere extension of high-quality video streaming duration may not necessarily yield an improvement in perceptual quality. Furthermore, the system is restricted to employing only one method, specifically buffer-aware, without considering additional observations
\par In ABRaider \cite{abraider}, Choi \textit{et al.} introduced
a multiphase RL approach that incorporates both online and offline phases to enhance adaptive video streaming. ABRaider integrates multiple ABR algorithms and generates suitable policies for different settings during the offline phase, with a focus on the specific users' environments during the online phase. Nevertheless, the methodology is complex and confusing due to the recent development of multiple ABR algorithms, leaving uncertainty over the necessary combination of these algorithms.  In addition, every algorithm possesses its own unique set of deficiencies, which are further compounded during the course of their execution and might negatively impact the overall efficiency. 
The work in \cite{wirelessai} devised a multimedia system comprising three fundamental components: end users, network, and servers.  Initially, they presented a DL model that includes data pre-processing, representation learning, and QoS/QoE prediction.  Subsequently, they introduced a mechanism for regulating the bit-rate using RL.
 In D-DASH \cite{d-dash}, Gadaleta \textit{et al.} introduced a Deep Q-Learning model that combines DL and RL techniques to enhance the QoE of DASH video streaming. They proposed and assessed multiple learning frameworks that integrate feed-forward and  RNNs using sophisticated techniques.  Nevertheless, these networks were incapable of managing jobs with continuous action spaces, as they usually kept and iterated the value functions of state-action pairs as a lookup table. Integrating the advantages of fuzzy logic with sophisticated DRL techniques, Yaqoob \textit{et al.} proposed FReD-ViQ \cite{fred}, an adaptive streaming solution that uses Fuzzy RL to provide excellent, personalized user experiences. Initially the fuzzy-logic models the network changes managing the vast dimensionality of the state space, which frequently obstructs learning-based algorithms. Then the  double Deep Q-Network is enhanced by integrating a Dueling structure, adaptive noise injection and a sampling approach employing prioritized experience replay. FReD-ViQ strikes a balance between exploration and exploitation, allowing for quick response to dynamic environments. It yields enhanced QoE performance, assessed through the instantaneous perceived quality of every segment, quality fluctuations amongst video segments, and the rebuffering occurrences.
\par  Table \ref{t:related_compare} presents a concise overview of the comparison of the recent ABR schemes. Table \ref{t:adaptive_streaming} summarizes the ML-based user-centric adaptive streaming approaches.
 \begin{table*}[!t]
 \scriptsize
 \centering
  \caption{Summary of ML-based user-centric adaptive streaming techniques}
   \label{t:adaptive_streaming}
\begin{tabular}{|p{10 mm}|p{21 mm}|p{56 mm}|p{21 mm}|p{18 mm}|p{10 mm}|}
\hline
Category     & ML Technique & Influence Factors & Accuracy & Datasets used& Reference paper\\ \hline
\multirow{7}{*}{\shortstack{Adaptive \\streaming}} &   Imitation learning (Neural Network)           &     past network, video, and playback features        &     7.5-16.79\% average QoE improvement  &  HSDPA\cite{hsdpa_url}, FCC\cite{fcc}, Oboe\cite{oboe}    &  comyco  \cite{comyco}  \\ \cline{2-6} 
 & Reinforcement Learning (policy gradient) &  network throughput, chunk download time, sizes of subsequent chunks, number of chunks, chunk bit-rate &       12-15\% QoE improvement  &   FCC, 3G/HSDPA       &     Pensieve  \cite{pensieve}   \\ \cline{2-6} 
&  policy gradient (online RL) &  throughput, download time, next segment size, buffer, \#remaining segments, previously requested segment size  &    2.5-28\% $\uparrow$ QoE than existing   &   FCC, 3G/HSDPA, Belgium\cite{belgium_4g}     &     ORL \cite{ictc2022}    \\ \cline{2-6} 
 &  Soft Actor-Critic (SAC) driven DRL (offpolicy)      &   throughput, download time, next segment size, buffer, \#remaining segments, previously requested segment size   &   27.42\% $\uparrow$ QoE than existing     &     FCC, Oboe, \cite{acmvideo}   &     SAC-ABR \cite{sac_abr}    \\ \cline{2-6} 
&  Deep Q-Learning &  previous downloaded segment quality, buffer, quality-rate of next segment, channel capacity  &    1.2-1.74 \%$\uparrow$ in SSIM (image quality)     &   Belgium       &     D-DASH  \cite{d-dash}    \\ \cline{2-6} 
&   DRL  (3D ConvNet)     &  bandwidth, buffer occupancy  &  9.3-12.9\%$\uparrow$ video quality than \cite{huang_bb, festive, mpc}                  &     FCC, 3G/HSDPA      &      \cite{wirelessai}   \\ \cline{2-6} 
 &  multiphase RL (online and offline) &   aggregation of ABR algorithms (BOLA, MPC, Festive etc)&  19.9\% (VoD), 42.2\% (live) QoE improvement     &  FCC,UCC\cite{ucc4g_lte}, UCC5G\cite{raca5g}, Belgium, HSDPA, Oboe     &     ABraider  \cite{abraider}     \\ \cline{2-6} 
  &  Fuzzy logic + Dueling Double Deep Q-Network &   Buffer occupancy, bandwidth, past bit-rates,  download time, segment count and segment
sizes &   QoE gain of 23.1\%, 23.97\%, 33.42\% than Linear, Log, and HD QoE models.    &   HSDPA \cite{hsdpa_url}    &     FReD-ViQ  \cite{fred}     \\ \hline
\end{tabular}
\end{table*}
\subsection{Intelligent streaming of 360$^\circ$ videos}
\par For 360$\degree$ video streaming, on tiling, segments with short duration are generated. The client can request tiles of varying quality levels in particular regions by fetching the relevant video segment files. In order to achieve flexible transmission, adaptive tiling techniques are used that not only improve responsiveness of the streaming system to  viewport variations, but also  boosts compression efficiency. In \cite{nguyen2018predictive}, based on the analysis that fixation point prediction error distribution is normal, Nguyen \textit{et al.} predicted user's future viewport. Then, a tile selecting algorithm determined the most recent viewport region repeatedly without latency and picked a set of tiles to transfer encompassing the region. Rossi \textit{et al.} \cite{rossi-navigation} designed an adaptive streaming system that considers tile-dependent coded content  and proposed an adaptation logic that can optimize the download rate of each tile on the client side based on user's navigational history. The optimization of tile-rate was formulated as an integer linear programming problem. 
\par Effective tiling methods need to balance trade-off between tile size selection and coding efficiency. First, tile size should be small, so that entire region of 360$^{\circ}$ views is covered without consuming additional pixels. On the contrary, tile size should not be too small such that the advantage of consuming lesser pixels is over-weighed by reduced coding efficiency. The second trade-off is between employing more tiles, which takes up more storage space but improves streaming efficiency against  using fewer tiles, which takes up less storage space but limits different tiling options during streaming. ClusTile \cite{clustile} approach picks a group of tiles that was encoded and saved on the server by resolving an integer linear programming  that was designed to approximately capture these trade-offs. However, the selected set of tiles it generates may overlap and lacks a straightforward method for choosing tiles from this set which reduces network bandwidth while streaming.
In \cite{zouprobabilistic}, Zou \textit{et al.} suggested a rate adaptation technique for users contending for server-side transmission capacity. The spherical viewport is mapped to planar projection based on the viewport forecast, and the visibility probability of every tile is calculated for each user. To reduce the experienced video distortion, the server chooses transmission tile-rate for each user. 
\par Ghosh \textit{et al.} \cite{ghoshrate} encoded the visible tiles in higher resolution and the rest in lower resolution depending on viewport data and fluctuating network constraints. They demonstrated that streaming at different quality levels for visible and non-visible regions can improve the efficacy of the QoE metric by $\sim$20\%. Xie \textit{et al.} \cite{xie_probdash} deployed a probabilistic strategy to prefetch tiles inorder  to reduce viewpoint estimation error and developed a viewpoint adaptive system. Pano \cite{pano} suggests a tiling method with tiles of different sizes to establish a good balance between perceptual video quality and efficiency with which it can be encoded. Dividing the projected video into tiles, Pano  aggregates smaller tiles into bigger ones for encoding, guaranteeing that regions with comparable content are encoded together, while those with substantially varied compressibility are encoded differently. 
\par In \cite{shafi_mtc}, Shafi \textit{et al.} decided the tiling strategy and resolution for streaming 360$^\circ$ video using multi-tile arrangements.
Within the adaptation interval, the size of the tiles were determined based on the vast circular distance between the original and forecasted viewports. Bit-rates are  subsequently assigned non-uniformly to the viewport and non-viewport areas. Graf \textit{et al.} \cite{graf_bw} compared the efficiency of various tiling patterns to that of conventional monolithic streaming. They analyzed that a $6\times4$ tiling pattern (compared to 8$\times$5, 5$\times$3, 3$\times$2, and 1$\times$1) would offer a beneficial compromise between coding efficiency and bandwidth usage for various categories of content.  Additionally, they demonstrated that they could achieve a considerable bandwidth reduction 
using a basic streaming method to stream a given viewport's data.
\par The upsampling models, specifically for super-resolution, have employed diverse networks such as  DNN \cite{srabr}, Deep CNN \cite{liu2022video}, Generative Adversarial Networks (GAN), and autoencoders. UNets \cite{unet360}, ResNets \cite{resnet360}, and DenseNets \cite{liu2022video} have been deployed in this field of work. Each of these models includes convolutional layers with several channels that can feed forward the residual information after some layers. A super-resolution embedded ABR method, SR-ABR \cite{srabr} incorporates content-induced performance variation of super-resolution DNNs into bit-rate decision making processes.   SR-ABR employs DRL to determine future bit-rates, catering diverse network conditions. To effectively leverage the content-induced performance variation of super-resolution-DNNs, the authors initially characterized this variability across diverse video content and subsequently employed a 2D convolution kernel to extract the features of performance variation of super-resolution-DNNs for a brief future video  chunk as one of their inputs. Super-resolution is an advantageous method to enhance the QoE for video streaming. 
\par In recent years, DRL has been used in bit-rate adaptation algorithms \cite{d-dash}, \cite{pensieve}. 
The DRL agent can acquire the bit-rate adaption policy through a series of exploration steps, with punishment and rewards determined by the actions chosen. In \cite{concolato}, ML was employed for video bit-rate adaption. Unlike the independent data samples in DL, the RL agents constitute strongly correlated states in the sequential phases and do not need hand-labelled data.
RL can be employed to determine the ideal strategy for selecting the quality \cite{chiarionline}. However, it is essential for the network to adhere to the Markovian Property. 
PENSIEVE \cite{pensieve} employs a RL technique for dynamic streaming, although it is specifically designed for conventional videos and is not well-suited for streaming tile-based 360$^{\circ}$ videos.
In \cite{hotdash}, the HotDASH  enhances video quality by pre-loading a segment with high temporal priority for streaming.
PARSEC \cite{dasari} adopts a different methodology that decreases the amount of bandwidth needed to transmit by utilizing a client-side computation approach with DNN model.  The prediction of FoV for 360$^\circ$ adaptive video streaming does not include the quality preference  for tiled streaming in \cite{dasari, xie_cls, sassa}. 
\par In \cite{srl360}, Fu \textit{et al.} proposed 360SRL, a Sequential RL (SRL) based ABR scheme for 360$^{\circ}$ videos. Initially, the decision space of 360SRL was transformed from exponential to linear by implementing a sequential ABR decision framework. Then, 360SRL generates ABR selections exclusively based on prior QoE performance, not precise bandwidth estimates. A similar approach using DRL is adopted in DRL360 \cite{drl360}, \cite{drlrate360}, \cite{rapt360}, and \cite{drl_jiang}. 
The DRL360 model contributes to system performance improvement by jointly maximizing several QoE criteria over a diverse collection of dynamic features. From the observations acquired by client side video players, the DRL360 model adaptively distributes rates for the tiles of next video frames.
In \cite{drlrate360}, Kan \textit{et al.} developed an Asynchronous Advantage Actor-Critic (A3C) algorithm for adapting the spatial and temporal rates using a constant tiling method in which the viewport estimation error was not considered. The authors in  \cite{rapt360} proposed RAPT360, a DRL approach for rate adaptation combined with adaptive prediction and tiling strategy for streaming 360$^{\circ}$ video. They designed a buffer occupancy-based viewport recognition approach to identify a user's viewport region, which could encompass the real viewport with any probable confidence level. They devised a viewport-aware adaptive tiling method to make better use of bandwidth. Jiang \textit{et al.} proposed RLVA \cite{drl_jiang}, RL-based Viewport-Adaptive 360$^{\circ}$ video streaming which improves  the model performance in viewport estimation, prefetch scheduling, and rate adjustment. They proposed a tile fetch scheduling algorithm to update tiles based on the most recent estimation results, reducing the negative impact of estimation error even further.
\par In \cite{li2023achieving}, the authors propose QBAM, a QoE-fairness bit-rate allocation algorithm to overcome the challenges arising out of difference in preference of individual users in choosing their bit-rates according to their viewports. Firstly, a clustering technique is used to examine preference of viewers. Then a DRL is used to train bit-rate allocation algorithm. Manfredi \textit{et al.} \cite{man2023lstm} presents LSTM-based approach for predicting viewports in immersive video systems. Through comprehensive evaluations, the LSTM-based viewport prediction system demonstrates superior performance in accurately anticipating user patterns, contributing to an improved overall immersive experience. It predicts user's point of interest by using temporal dependencies within user interactions, making the response more responsive and immersive. In \cite{islam2023}, Islam \textit{et al}. creates ML models that can be used by network operators to regulate real-time QoE of virtual reality video sessions. In their proposed solution, packet level information is used as data input for training the ML model to estimate target QoE. Through experimental test results in 4G and 5G environment, the study establishes that the trained model provides output of reasonable accuracy and successfully predicts QoE for HTTPS and QUIC. The study in \cite{zhang2018saliency} presents content-driven viewport predictor framework that is integrated with personalized federated learning methods and data fusion techniques. First, a saliency detection model is framed, which relies on a Spherical Convolutional Neural Network (SPCNN) for extracting salient regions from 360$\degree$ video frames. Then an algorithm is presented for head movement prediction which incorporates and increases the accuracy using personalized federated learning. Finally, the viewport prediction framework is designed, whereby the fusion approaches are used to generate a fused feature map out of the saliency map and head orientation map. To tackle the issues of faulty network estimations and inherent saliency bias that restricts enhancements in QoE, Wang \textit{et al.} developed a resilient saliency-driven quality adaptation framework for 360$\degree$ video streaming, RoSal360 \cite{robust_saliency}. They designed the model for predicting the transmission duration of video tiles using a decoupled self-attention architecture and a DNN that considers the tile size. Also, they developed an online correction approach that is driven by RL to effectively compensate for the incorrect quality allocations caused by saliency bias. Through comprehensive prototype trials conducted on actual wireless networks, RoSal360 enhances video quality as well as minimizes rebuffering.
\par FBRA360 \cite{fbra360} presents Fuzzy-Based bit-rate Adaptation Scheme to deal with viewport prediction and bit-rate selection demanding situations in tile-based streaming. It uses a fuzzy based controller to control weighted mix of viewport from each user's history trajectory and multi-user attention distribution along with the buffer occupancy. In UVPFL \cite{federated360}, the viewport is predicted using Federated Learning based on user-profiles. Federated Learning is used to forecast the user's viewport by analysing their head movement patterns across various video categories. It utilises both the observed client head movements and the past viewport data to make predictions.  The obtained viewports are then compared with either a similar user viewport or a historical viewport. The predicted viewport is then merged and updated if overlapping area exceeds a certain value. MOSAIC \cite{mosaic} presents an end-to-end video streaming system that predicts viewport and delivers tiles encoded at suitable bit-rates to optimize end user video quality within the network capacity. To forecast viewports, a CNN model integrated with LSTM is trained using motion, saliency maps, and client's head tracking data that learns both spatial and temporal features individually. Further,  3DCNN is used to capture the spatio-temporal features. DeepVR \cite{deepvr} provides a tile-based adaptive immersive video streaming system with DRL-agent that can effectively adjust its bit-rate allocation strategy according to the changing environment. 
The FoV in the following seconds is predicted using attentive LSTM networks. DeepVR obtains a better QoE score through the incorporated DRL algorithm Rainbow, outperforming current panoramic video streaming systems. NOVA \cite{nova} is an effective Neural-Optimized Viewport Adaptive streaming system designed to enhance the QoE of users. In NOVA, initially a foveated rendering super-resolution method executes super-resolution operations at the tile level to convert low into high-resolution video tiles at the edge server. Then a meta-learning-based multi-agent RL algorithm is proposed to effectively learn video tile selection and super-resolution enhancement decisions to optimize the long-term user experience under prevailing network constraints. Comprehensive experimental findings indicate that NOVA significantly enhances average QoE while utilizing less bandwidth than existing methods.
\par In \cite{vaser360}, a 360$\degree$ live video ingest system, Vaser takes into account viewport information for neural enhancement and tile uploading. It considers viewport data to cut down on upload bandwidth requirements without compromising user experiences. 
It proposes an improved patch selection method to increase the super resolution model training efficiency and taps into DRL for setting upload stream bit-rate and super resolution model update frequency. The study in \cite{da2018} introduces a two-stage approach to forecast the quality of virtual reality videos streamed over mobile networks. Initially, they predict the video playout performance metrics employing regression trees using network QoS indicators and video structure as input. Subsequently the obtained playout metrics are used to model and estimate the experienced video quality. The work in \cite{fei2019} proposes a subjective and objective framework for virtual reality QoE evaluation that can meet the requirement of real applications. The subjective evaluation component utilises four dimensions, i.e., the following scores: quality, immersion, non-spinning sensation and global score. The objective component employs an enhanced neural network, which was created by incorporating the characteristics of psychology and cognitive neurology. 
\par The study in \cite{wu2020spherical} proposes a spherical-CNN model designed for accurately predicting the long-term viewport in 360$\degree$ videos. The network employs feature extraction to condense spatial-temporal 360 information. The work in \cite{feng2019exploring} attempts to address the stringent criteria for video processing time as the delay caused by viewport prediction adds to live streaming latency. Also, it tries to avoid user video trails and user traces, as these data might not be available to construct the prediction model. So, they change the workflow of CNN (AlexNet) model and training/testing process to achieve this. Peng \textit{et al.} \cite{spherical360} offers a DNN model based on spherical convolution to learn spatial aspects of 360◦ videos. The approach encodes distortion invariance into the CNN architecture. MAIVS \cite{maivs} presents an adaptive UHD 360$\degree$ immersive video streaming solution that leverages machine learning techniques. The solution aims to mitigate the data-rate demand associated with streaming high-resolution (UHD) videos. The process involves spatially dividing the 360$\degree$ videos into MCT. After being downscaled, these tiles are then encoded using HEVC. A DNN is trained to enhance the resolution (at the client) of the encoded tiles by upscaling it. The encoded tiles, along with the model parameters, are packaged into mp4 containers at various quality levels. The DASH technique is employed to stream video tiles and model parameters progressively. The Deep Q-Network (DQN) is trained using feedback (like, video quality parameter, buffer and network conditions) to carefully select the bit-rate quality segments. 
\par The survey in \cite{mahmoud_survey} explores how to best deliver 360$\degree$ videos on mobile devices, with a focus on edge caching and multicasting. A variety of caching namely tile based caching, collaborative caching, proactive and projection based caching can bring in optimization in streaming. Reinforcement based dynamic caching using ML is the most sophisticated of all. Furthermore, the research delves into the possible advantages of multicasting, which include lowering network latency and enhancing scalability while effectively delivering 360\degree videos to numerous users at once. In \cite{shot2023}, Prabavathy \textit{et al.}  presents a real-time shot boundary detection system for live 360$\degree$ virtual reality streaming using DL. The system automatically identifies transitions between shots, enhancing the seamless viewing experience in live VR environments. Its real-time capabilities ensure smooth adaptation to content that changes based on data and preferences, intensify the overall viewing experience. An innovative approach in \cite{omniscent} transforms the virtual reality experience by fusing 360° video streaming with omnidirectional olfaction technology. The immersive quality of 360° videos is improved through the seamless integration of scent delivery with both visual and audio stimuli. The system aims to enhance overall sensory engagement and capture viewers. Scents are discharged in the same direction as relevant objects appear. 
 \begin{table*}[!t]
  \scriptsize
 \centering
  \caption{Summary of ML-based 360$\degree$ adaptive video streaming}
   \label{t:360_streaming}
 \centering
\begin{tabular}{|p{8 mm}|p{14 mm}|p{33 mm}|p{12 mm}|p{1.6 in}|p{23mm}|p{8 mm}|}
\hline
\shortstack{Categ-\\ory}     & ML Technique & Influence Factors & Datasets used & Major observation/ application  & Accuracy & \shortstack{Referen-\\ce paper}\\ \hline
\multirow{16}{*}{\shortstack{360$\degree$ \\video\\ stream-\\ing}} &   Deep Q-Learning      &   network throughput, buffer status, segment PSNR  & \cite{hmd_corbi}  &    mitigate data-rate demand for streaming UHD 360\degree videos      &    8.52\%,48.18\% lowered bit-rate segments than PARSEC\cite{dasari} \&  \cite{pensieve}  &  \cite{maivs}   \\ \cline{2-7}
 & multi-agent RL & user viewpoint trajectory \& preference for video quality, buffer time, quality switching & 3G/HSDPA, \cite{wu2017} & obtaining QoE equitable bit-rate allocation for 360° video transmission & quality difference between users is 6-7  &  \cite{li2023achieving} \\ \cline{2-7}
 & LSTM & time samples,viewport trajectory,playout buffer & Avtrack \cite{avtrack},\cite{ag2019,nas2019,rossi2020,explor2017} & addresses critical challenges of viewport-adaptive streaming, showcasing efficacy of LSTM & prediction accuracy $\sim$60\% &  \cite{man2023lstm} \\ \cline{2-7}
& Tree-based methods, DNN \& K-NN & Throughput, \#packet interarrival time, bit-rate of streaming, stall time, startup delay, quality switch & \cite{islam2023data} & predict QoE using encrypted network level QoS data & RMSE of 0.09  & \cite{islam2023}\\ \cline{2-7}
& spherical CNN+ Federated Learning & video frames, head movement measures& \cite{zhang2018saliency} & content-driven viewport prediction based on saliency  \& head orientation maps & 4.35\%, 5.88\%, 7.46\%$\uparrow$ than existing  & \cite{set2023con} \\ \cline{2-7}
 & fuzzy logic controller & Buffer occupancy, throughput deviation, user behaviour bias & SalientVR \cite{salientvr}, \cite{band4g}  & Establishes bidirectional dependency between bit-rate selection \& viewport prediction & video quality 7.5\%$\uparrow$ than existing  & \cite{fbra360}\\ \cline{2-7}
  & Sphere U-Net, LSTM & spatial features, user's trajectory, saliency map &\cite{zhang2018saliency} & saliency network embedded with FoV prediction framework for better prediction  & 33-71\% relative LCC$\uparrow$ in saliency detection & \cite{spherical360} \\ \cline{2-7}
& Federated Learning & Tiles of current frame, saliency data & \cite{lo2017} & Improves viewport prediction in absence of historical data & 90\% (viewport prediction), 1.12- 64.9\%$\uparrow$ than existing & \cite{federated360}\\ \cline{2-7}
 & Deep Learning & viewport info, throughput, ratio of received frames, SR gain & \cite{xu2018gaze, ntire},\cite{mahimahi}, FCC & reduces 360$\degree$ live video upload bandwidth needs, without affecting quality & bandwidth need 40-55.6\%$\downarrow$, utility 1.15-3.61$\times \uparrow$   & \cite{vaser360} \\ \cline{2-7}
 & DRL& Throughput, remaining chunks \& tiles, rebuffering time, next tile size \& viewing probability, bit-rate & \cite{hsdpa_url},FCC, \cite{belgium_4g},\cite{lo2017} & optimizes streaming in viewport prediction, prefetch scheduling, \& rate adaption & 4.8-66.8\% $\uparrow$ in QoE than existing & \cite{drl_jiang} \\ \cline{2-7}
& CNN (ResNet 101)+LSTM  \& 3DCNN & saliency, motion map, head tracking trace &\cite{lo2017}, 4G/LTE \cite{belgium_4g} & combines neural network-based viewport prediction \& tile-rate allocation, optimizing video quality & 47-191\% $\uparrow$ video quality than existing & \cite{mosaic} \\ \cline{2-7}
& Decision Trees & delay, packet loss, throughput, startup delay, quality, quality switches count, video stalls & \cite{wu2017}, \cite{belgium_4g} & predicts perceived quality using network QoS indicators \& video playout performance metrics & predicts perceived quality with 4\% $\downarrow$ prediction error & \cite{da2018} \\ \cline{2-7}
 & Improved Neural Network (INN) & bandwidth, latency, packet loss, quality score; immersion, non-spinning \& global score & own \cite{fei2019} & predicts quality using no-reference two-stage improved neural network &  8.9\%$\uparrow$ in PLCC than traditional NN   & \cite{fei2019} \\ \cline{2-7}
& spherical CNN & user's viewing history, head motion info & - & long-term viewport prediction using spherical-CNN & 40-46\% viewport prediction accuracy   & \cite{wu2020spherical} \\ \cline{2-7}
& CNN (AlexNet) & Tiles, head movement traces & \cite{wu2017} &  viewport prediction accuracy with low bandwidth \& low timing overhead & 70-88\% prediction accuracy & \cite{feng2019exploring}  \\ \cline{2-7}
 & attentive LSTM, DQN &  viewpoint trajectory, past throughput, chunk size, buffer, bit-rate, remaining chunks, download time & \cite{david2018data}, \cite{hsdpa_url},  \cite{ucc4g_lte},  \cite{speedtest} & Tile-based adaptive streaming with DRL policy adjustment & 16\%$\uparrow$ in QoE than existing  &  \cite{deepvr}  \\ \cline{2-7}
 & Multi-agent RL  & downlink throughputs, tile dimensions for prior tile downloads, download durations, viewport trajectories & \cite{belgium_4g}, \cite{lumos5g}, \cite{drl360} &  optimized tile-selection and edge-assisted super resolution enhancement for maximizing QoE and adapting to network changes & enhances user-perceived QoE up to 27\% & \cite{nova}  \\ \cline{2-7}
  & DNN+RL  & tile throughput, tile size, saliency-based behavior prediction\! \& \!viewport prediction accuracy, buffer occupancy,time before playback for tile,tile’s TTP result,TTP
confidence & \cite{salientvr_url}, \cite{xu2018gaze} & saliency-based video quality adaptation framework that incorporates 
tile-size-aware transmission time estimation DNN model and a saliency-aware quality distribution for mobile video streaming  & enhances QoE (31.45\% MOS gain on average) & \cite{robust_saliency} \\ \hline
\end{tabular}
\end{table*}
\par The study in \cite{combinedfov} presents an adaptive streaming method that combines two FoV prediction techniques to provide more accurate dynamic viewing area recognition, allowing for interactive tile choices. It uses priority-based bit-rate adaptation approach that enhances end-user QoE. \cite{set2023con} presents a content-based viewport predictor framework, integrated with personalized federated learning (PFL) methods and data fusion techniques. A  saliency detection model is presented, which relies on a spherical convolutional neural network (SPCNN) for extracting salient regions from 360° video. An algorithm is presented for head movement prediction which incorporates and increases the prediction accuracy using PFL. Then, the viewport prediction framework is discussed, whereby the fusion approaches are used to generate a fused feature map out of the saliency map and head orientation map. Finally, the proposed framework is evaluated in terms of accuracy and precision metrics against existing viewport prediction algorithms.
\par Table \ref{t:360_streaming} summarizes the ML-based 360$\degree$ adaptive video streaming techniques along with IFs, major observations and used datasets.
\section{Video Quality Datasets and Performance metrics}
\label{s:data_per}
\subsection{Subjective video quality datasets}
This subsection discusses the commonly used video quality  databases such as LIVE \cite{d12sheikh2005live}, VQEG HD3 \cite{vqeghd3}, VQEG HD4 \cite{vqeghd3}, LIVE Netflix \cite{d16livenetflix1another}, LIVE NFLX II \cite{d59sheikh2005live3}, Waterloo \cite{d17wdb}, CSIQ \cite{csiqvqa}, {LIVE HTTP video streaming} \cite{chen}, LIVE Avaasi \cite{avaasi}, and LFOVIA \cite{lfovia}. 
The video sequences in the datasets contain various natural scenes, movie clips,  sports, animation, music, advertisements, documentaries etc.
The databases  cover a diverse set of videos with a broad range of Spatial Information (SI) and TI values. Table \ref{t:database} lists the frequently used video quality databases with various distortions, video characteristics, and display device .
 \begin{table*}[!t]
 \centering
\caption{Subjective Databases Overview. SV: Source Videos, DV: Distorted Videos, TE: Transmission error, cont.: continuous}
 \label{t:database}
 \setlength{\tabcolsep}{9pt}
\begin{tabular}{|p{52pt}|p{28pt}|p{7pt}|p{14pt}|p{27pt}|p{42pt}|p{63pt}|p{35pt}|p{25pt}|}
 \hline \smaller{Database} & \scalebox{0.79}{Resolution} &\scalebox{0.7}{\#SV} & \smaller{\#DV} & \smaller{Duration (sec)} & \smaller{Scores range} & \smaller{Distortion Type} & \smaller{Display Device} &  \smaller{Rating type} \\
 \hline 
\smaller{LIVE \cite{d12sheikh2005live}} & \smaller{768$\times$432}&\smaller{10} &\smaller{150}   &   \smaller{10} & \smaller{[0,100]}&  \smaller{H.264, MPEG 2,} TE & \smaller{CRT Monitor} & \smaller{overall}\\  \hline  
\smaller{VQEG HD3 \cite{vqeghd3}} & \smaller{1080p} &\smaller{13} & \smaller{155}  & \smaller{10} & \smaller{{[0,5]}}
 & \smaller{H.264, MPEG 2,} TE &  \smaller{Samsung}  \scalebox{0.7}{TV} & \smaller{overall}\\  \hline
 \smaller{VQEG HD4} & \smaller{1080i} &\smaller{13} & \smaller{155}  & \smaller{10} & \smaller{{[0,5]}}
 & \smaller{H.264, MPEG 2,} TE & \smaller{LG} \smaller{TV} & \smaller{overall}\\  \hline
 \scalebox{0.7}{LIVE Netflix \cite{nflx1}} & \smaller{1080p} & \smaller{7} & \smaller{112}  &  \smaller{$>$60} & \smaller{[0.38,4.97]} 
 & \smaller{H.264,scaling, freezing} & \smaller{Mobile} & \smaller{cont.}\\ \hline
\scalebox{0.7}{LIVE NFLX II \cite{d59sheikh2005live3}} & \smaller{1080p}& \smaller{15} & \smaller{420}  &  \smaller{$\geq$25} & \smaller{[0.19,4.9]} & \smaller{H.264,scaling, freezing} & \smaller{Computer monitor} & \smaller{cont.}\\ \hline 
\smaller{Waterloo \cite{d17wdb}} & \smaller{1080p}& \smaller{20} & \smaller{180}  & \smaller{$\approx 15$} & \smaller{[0,100]} & \smaller{compression,TE} & \smaller{LCD Monitor} & \smaller{overall} \\  \hline 
\smaller{CSIQ \cite{csiqvqa}}  & \smaller{832$\times$480} & \smaller{12} & \smaller{216} & \smaller{$10$} & \smaller{[0,100]} & \smaller{compression, TE}& \smaller{LCD Monitor} & \smaller{overall}\\  \hline 
\smaller{LFOVIA \cite{lfovia}}  & \smaller{2K, 4K} & \smaller{18} & \smaller{36} & \smaller{$120$} & \smaller{[0,100]} & \smaller{encoding,freezing} & \smaller{CES Android app} & \smaller{overall+ cont.}\\  \hline 
\smaller{Mobile stall II \cite{mobilestall2}}  & \smaller{1280$\times$720, 1280$\times$640} & \smaller{24} & \smaller{174} & \smaller{29-134} & \smaller{[19.12, 75.82]} & \smaller{freezing} & \smaller{Mobile} & \smaller{overall+ cont.}\\  \hline 
\smaller{LIVE Avaasi Mobile}\cite{avaasi}  & \smaller{1280$\times$720, 640$\times$360} & \smaller{24} & \smaller{180} & \smaller{29-134} & \smaller{[0,3]} & \smaller{stalling+startup delays} & \smaller{Apple iPhone} & \smaller{overall}\\  \hline 
\scalebox{0.7}{LIVE HTTP video} \scalebox{0.8}{streaming \cite{chen} } & \smaller{720p} & \smaller{3} & \smaller{15} & \smaller{300} & \smaller{[30,70]} & \smaller{quality switches} & \smaller{TV} & \smaller{overall+ cont.}\\  \hline
\smaller{MCQoE} \smaller{\cite{mcqoe}}  & \smaller{360p-2160p} & \smaller{7} & \smaller{14} & \smaller{60} & \smaller{[0,100]} & \smaller{H.264, TE, freezing} & \smaller{TV,PC,HD- phone} & \smaller{overall+ cont.}\\  \hline 
 \end{tabular}
 \end{table*}
\par 
The damaged video sequences for every source video in the LIVE dataset \cite{d12sheikh2005live} are acquired using the following techniques:
i) There are 4 test videos that use H.264 compression. The compression rates for these videos range from 200 Kbps to 5 Mbps. Additionally, there are test videos that use 'MPEG-2' compression with rates ranging from 700 Kbps to 4 Mbps. 
ii) There are three IP test videos (using RTP) created by intentionally dropping packets in a particular error pattern specified by the Video Coding Experts Group (VCEG) and have loss rates of 3\%, 5\%, 10\%, and 20\% from the H.264 compressed videos.
 iii) There are four test videos created by transmitting H.264 compressed videos over error-prone wireless networks. The errors in these networks are simulated using bit error patterns resulting in packet error rates ranging from 0.5\% to 10\%.
\par The videos in VQEG HD3 \cite{vqeghd3} are subjected to compression artifacts caused by differences in framerate, bit-rate, and codec type, as well as transmission artifacts.
Transmission error in IP networks (via User Datagram Protocol (UDP)) encompass packet losses that occur randomly, bit errors, and variations in packet delay. The packet loss rates (PLR) are 0.015\%, 0.024\%, 0.035\%, and 0.3\%. The VQEG HD4 dataset \cite{vqeghd3} includes similar impairments such as compression and transmission artifacts, as well as varying levels of packet losses. However, VQEG HD4 had varying viewing conditions, including differences in monitor size, dot pitch, calibration process, refresh rate, and bit depth, detailed in \cite{vqeghd3}.
\par In \cite{nflx1}, the LIVE Netflix dataset consists of streaming videos that contain Netflix content and video sequences. The distortions in these videos are caused by H.264 compression, stalling, and a mix of both. They created eight playout patterns, with one of these patterns assuming sufficient bandwidth to allow the client to play out content at a consistent rate (assuming no network impairments). Rest of the patterns replicate severe impairments by taking into account the depletion of the buffer. It considers both impairment free as well as impaired network conditions.
\par In LIVE NFLX II \cite{d59sheikh2005live3}, the streaming scenario was modeled by four main modules: encoding, video quality, network transmission, and client. The encoding module created encodes by determining the encoding parameters, like resolution and QP, using an optimization framework. The video quality module assessed the quality of the video. The network transmission module accounted for the effects of varying network conditions. The client module requested the next video chunk to be played using four ABR algorithms.  In order to quantify the effects of varying network circumstances, a selection of seven network traces was made from the HSDPA dataset, which contains authentic 3G traces collected during various travel journeys throughout Norway. The traces encompass network behaviors that range from low to high bandwidth conditions, which can lead to rapid fluctuations in bit-rate or quality and buffering. ABR algorithms effectively measure the influence of adjustments on the client side QoE, a factor that is absent in other datasets.
\par
The Waterloo dataset \cite{d17wdb} includes streaming videos' quality, where the videos are encoded using the H.264 codec and have three various bit-rates (i.e., 500, 1500, and 3000 Kbps) to achieve varying quality levels. Stalling events, each lasting five seconds, were added at the beginning or middle of these encoded sequences. 
In CSIQ dataset \cite{csiqvqa} there are a total of eighteen categories that have been distorted, and within these categories, there are six different kinds of distortions. The distortion kinds include four video compression methods (H.264, Motion JPEG, HEVC,  and wavelet compression) and two transmission-based methods (e.g., packet loss in a wireless network and additive white Gaussian noise).  It
encompasses a variety of framerates, including 24, 25, 30, 50, and 60 frames per second (fps). 
\par The LFOVIA dataset \cite{lfovia} comprises nine source videos, each in Full HD and Ultra HD resolution. The collection of 36 videos exhibits distortions that result from rate adaption and rebuffering. The rebuffering frequency ranges from 0.5 to 5 occurrences per minute. Bit-rate adjustments are made both upward and downward, wherein the video rate is altered to a higher or lower level.
The subjective scores obtained from LIVE and CSIQ was in the form of DMOS, whereas rest of the databases were in terms of MOS. The subjective scores in various databases are in different ranges. 
\par The Mobile stall II \cite{mobilestall2} comprises a collection of 174 streaming videos on mobile devices, featuring 26 distinct stalling patterns. The patterns include duration, position, and recurrence rate of stall occurrences. It contains videos with quality varying over time during the playback. It has both per-frame continuous as well as overall QoE scores.
\begin{table*}[!t]
\scriptsize
\centering
\caption{Overview of databases for different network conditions}
 \label{t:traces}
\begin{tabular}{|p{15 mm}|p{3.7 in}|p{13 mm}|p{11 mm}|}
\hline
Dataset    & Description & Network & Download url \\ \hline
FCC  \cite{fcc}& {Raw data collected from fixed ISPs in US. It includes over 1 million throughput measures with average throughput at 5-second intervals.} &  Broadband  &  \cite{fcc} \\ \hline
HSDPA  \cite{hsdpa_url}& {HDSPA gathers 30-minute uninterrupted video streaming throughput data from mobile devices throughout several Norwegian routes utilising various modes of transportation (e.g., vehicle, bus, train)} &   3G  & \cite{hsdpa_url}\\ \hline
UCC \cite{ucc4g_lte} & {It includes 4G traces of client-side cellular KPIs from two Irish mobile operators, covering various mobility patterns.  It contains 135 traces with an average time of 15 minutes and observable throughput from 0 to 173 Mbit/s at one sample per second.} &    LTE  & \cite{ucc4g_lte_url} \\ \hline
UCC5G \cite{raca5g}     & {It contains 5G traces gathered from Irish mobile operator, that includes static and mobility patterns for video streaming and file download. It comprises of client-side cellular KPIs  consisting of  channel, context, cell-related metrics and throughput information} &    5G/ LTE  & \cite{raca5g_url} \\ \hline
Belgium \cite{belgium_4g}   & {
This report includes 40 throughput traces for 4G networks in Ghent, Belgium, collected using 6 transportation modes  over 5 hours of monitoring. The bandwidth ranged from 0 to 111 Mbps, averaging 30.3 Mbps $\pm$ 16.7 Mbps}
 &   4G/ LTE & \cite{belgium_4g_url} \\ \hline
Oboe trace \cite{oboe} & {The throughput traces contain chunk sizes and their download times for on-demand video sessions. Traces include desktop sessions  with wired and mobile device  sessions with WiFi or cellular connections.  Around 5K traces were collected from wired PCs and 4K traces from WiFi or 3G/4G mobile devices.}
& WiFi/3G/ LTE  & \cite{oboe_url} \\ \hline
E2E dataset of video streaming \cite{e2e} & It presents a variety of comprehensive metrics, referred as Key Quality Indicators (e.g., number, frequency, duration of stalls) for evaluating the end-to-end performance of video streaming and cloud gaming  across various network technologies  & LTE/5G/ Ethernet/ WiFi &     
\cite{e2e_url}  
\\ \hline
\end{tabular}
\end{table*}
\par The LIVE-avaasi \cite{avaasi} is a mobile video dataset that simulates network impairment distortions which includes stalling events and start-up delays. It replicates realistic stalling patterns by altering multiple QoE-influencing factors, including the location, frequency, and duration of the stalls as well as the kind of video content on end users' QoE. It has 24 source videos with 180 distorted videos having 26 distinct stalling patterns and 4830 opinions gathered from 54 individuals.
\par The LIVE HTTP \cite{chen} dataset facilitates the creation of TVSQ prediction models for HTTP-based video streaming.  To mimic the quality changes observed in HTTP-based streaming, the STSQs of the videos were created to fluctuate randomly across time scales spanning multiple seconds. It has 15 videos, each 5 minutes long, and viewed by 25 participants.
\par The MCQoE \cite{mcqoe} is a Multi-device Continuous QoE dataset which is publicly available. It has over 76,000 continuous and 1260 overall scores collected from 60
participants. It incorporates artificial glitches, like rebuffering events and fluctuations in quality, to replicate the video streaming experience. The ratings are collected across several devices, i.e.,  75-inch UHD TV, 24-inch
QHD PC, and 6.1-inch HD smartphone.
\par Table \ref{t:traces} summarizes the  publicly available databases that cater to various network conditions.
\begin{table*}[!t]
\scriptsize
\centering
\caption{Overview of databases for 360$\degree$ videos containing records of head movements, eye tracking and viewport traces.  FR/SF: framerate/ sampling frequency}
 \label{t:hmd_data}
\begin{tabular}{|p{7 mm}|p{43 mm}|p{11 mm}|p{6 mm}|p{11mm}|p{8 mm}| p{8 mm}|p{22 mm}|p{10 mm}|}
\hline
Dataset    & Description & Subjects & No. of videos& Resolu- tion &FR/SF& duration (sec.) & Watching device & Downlo- ad url \\ \hline
\cite{hmd_corbi}  &  head movement of users 
& 59 (48 males, 11 females) & 5 & 3840×2048 & 25, 30, 60 fps& 70  & Razer OSVR HDK2 HMD &  \cite{hmd_corbi_url} \\ \hline
 \cite{wu2017} 
 & head movement of users, records of user behavior & 48 (24 males, 24 females) & 18& 2560x1440& 25, 29, 30 fps &146-506 & HTC Vive headset & \cite{wu2017_url}\\ \hline
Avtrack \cite{avtrack}  &   HMD's current tracking location, current head rotation, video playback time along with timestamp.  & 48 (25 female, 23 male) & 20 &4K&-&30  & HTC Vive HMD & \cite{avtrack_url}\\ \hline
 \cite{ag2019}  & eye-tracking recordings i) Gaze point's x and y coordinates ii) head direction coordinates and its tilt. & 13 &14& upto 4K&120 Hz & 60  & FOVE VR headset+ eye tracker& \cite{ag2019_url}\\ \hline
 \cite{nas2019}  & viewport traces and their heatmap, response  by the subjects to questions& 60 (17 female, 43 male)& 28 & upto 3840x2160 & 29, 30 fps&60  & Oculus Go HMD & \cite{nas2019_url}\\ \hline
 \cite{rossi2020}& navigation trajectories using several viewing platforms & 94 (29 females, 65 males)&15&  2560×1440 &24, 25 30 fps & 20 
& Oculus Rift HMD, Alienware15 Laptop, Apple iPad Pro10.5 tablet & \cite{rossi2020_url}\\ \hline
 \cite{zhang2018saliency} & ground truth (original) gaze points of the observers &27 & 104& upto 4K &- & 20-60 & HTC Vive HMD+ 7invensun a-Glass eye tracker & \cite{zhang2018saliency_url}\\ \hline
\cite{salientvr}   &  gaze-annotated dataset that would help in attention behaviour analysis. &30 (14  female, 16 male) &20 &4K& 30-60 fps& 120-660 & HTC Vive Pro Eye VR headset&  \cite{salientvr_url}\\ \hline
 \cite{xu2018gaze}  & eye tracking dataset &45(20 females,25 males)& 208 &4K& 25 fps&20-60 & HTC Vive HMD+ 7invensun a-Glass eye tracker & \cite{xu2018gaze_url}\\ \hline
\cite{dharma}& more than 3700 viewport traces&- &88& upto 4K&10 Hz & 30-655  & HMD& \cite{dharma_url}\\ \hline
\end{tabular}
\end{table*}
\subsection{HMD, Eye-tracking and Viewport traces Datasets} \label{s:per_metrics}
In \cite{hmd_corbi}, the dataset contains head movement of users recorded from 59 people while watching 360$\degree$ videos each of duration 70 sec.  The chosen videos cover a broad range of 360$\degree$ information, thus varying user engagement and navigation patterns. The user's head position log files contain the timestamp, frame ID, and unit quaternion values that are used to determine the user's head position.
The dataset in \cite{wu2017} has data records from 48 people that viewed eighteen 360$\degree$ videos across 5 categories. The way viewers watch the videos, their head movement during each session, where they focus, and what they can recall are all recorded following every session. The log files comprises of the timestamp, playback time, unit quaternion values and position of the HMD device. It contains demographic profiles and records of user behavior.
\par Avtrack \cite{avtrack} presents visual assessment of 48 participants who viewed 20 distinct and interesting 360$\degree$ videos.  As the subjects watched the contents, their head movements were captured. It gives the participants' exploration activity by providing the angular ranges they covered and an analysis of the specific regions where they spent most of their time. The gathered data can also be displayed as head-saliency maps. The available data includes HMD's current tracking location, current head rotation, video playback time along with timestamp.
In \cite{ag2019}, the dataset provides an eye-tracking recordings for both real-world 360$\degree$ videos and a synthetic video clip. The recordings have a comparatively high frequency of 120 Hz, making it easier to infer eye movements. Details include i) Gaze point's x and y coordinates (in equi-rectangular coordinate) throughout the 360$\degree$ video surface ii) The head direction coordinates and its tilt. 
\par The dataset in \cite{nas2019} comprises viewport traces collected from 60 people  who watched the 360$\degree$ videos. Furthermore, it offers feedback of the viewers on their experience following the viewing of each video. The log files include timestamp, the quaternion components representing rotation of the viewport, and cartesian coordinates of the vector pointing towards the center of the user's viewport. The dataset also includes response given by the participants to the questions, heatmap of viewport traces, and pitch and yaw angle histograms for every video.  In \cite{rossi2020}, it offers navigation trajectories obtained for heterogeneous 360$\degree$ videos using several viewing platforms.  The subjective experiments were conducted at Trinity College Dublin and University College London. It examines the behavior of users across different content and viewing devices. Initially, analysis is conducted using metrics like angular velocity and viewport center distribution that emphasizes the extent to which the user's navigation is influenced by display device. Next, the similarity among users is examined based on the viewport that is displayed over a period of time.
\par In \cite{zhang2018saliency}, the dataset comprises of the ground truth (original) gaze points of the observers, collected  from more than 20 subjects for 104 video clips. The 360$\degree$ source videos are collected from Sports-360 dataset (contents like basketball, skateboarding, parkour, BMX, and dance) that have  duration  of 20 to 60 sec. The HMD has an integrated eye tracker that records the participants' eye fixation positions while they watched the videos. The dataset in \cite{salientvr} presents a gaze-annotated long 360$\degree$ video dataset that would help in attention behaviour analysis. The dataset is further processed as follows for more practical applications: i) The missing values (produced by eye closure) are handled using moving average interpolation ii) data is transformed from millisecond level into frame level iii) the head directions' quaternion coordinates are changed to 2D plane coordinates.
\par Xu \textit{et al.} \cite{xu2018gaze} presents an  extensive eye tracking dataset for dynamic VR scenes. There are 208 HD 360$\degree$ videos (resolution-4K, 25 fps) of duration 20-60 sec. in the collection, and at least 31 participants as viewers. The dataset analysis reveals that contents of the image and history of the scan path influence gaze prediction. It includes a wide range of video content, including , including indoor and outdoor scenes, music, sports etc. A combined dataset in \cite{dharma} consists of 88 videos of duration $\geq 30$ sec. and includes more than 3700 viewport traces, which contain a total of more than 142 hours of watching records. Various formats of distinct datasets are merged such as head orientation, trace duration, and data sampling rates. Yaw and Pitch angles were expressed in radians and roll angle changes were ignored. To address the disparities in sampling rates, the complete dataset was re-sampled at 10 Hz. 
\par  Table  \ref{t:hmd_data} lists the various publicly available datasets for 360$\degree$ videos containing head movement directions, viewport traces, and eye tracking records.
\section{Open Challenges}
\label{s:open_challenge}
User-centric multimedia streaming systems have played a pivotal role in enabling convenient access to multimedia content, while improving the service quality. Although several issues pertaining to it have been addressed, still there are some issues that need to be considered for future research.\\
1) \textit{Subjective Quality Assessment:} The developed subjective databases used as ground truth to predict the video quality ratings have design limitations. Conventional subjective video quality databases are usually created by initially choosing a limited number of excellent sources, known as ``reference", and subsequently artificially altering them using simulated methods. These databases have little variation in their content characteristics and distortions, which makes them unable to accurately replicate complex impairments in real-world user created videos. All visual media share fundamental principles of perception. However, to comprehend artifacts and distortions specific to a given domain, it is necessary to create subjective databases specifically designed for each domain. Constructing  extensive, diversified, and impartial subjective video datasets is still a crucial area of study for researchers. Additionally, because of the small size, the publicly available subjective databases devoted to video quality are unable to fully utilize the potential of DL techniques. 
\par Immersion is a relatively new QoE attribute. Currently, immersion does not  have a standardized quantitative quality measurement index. Various IFs and the intricate interplay between them could be contributing factors. The available subjective assessments primarily address visual quality, presence perception, physical comfort, and multisensory integration. Subjective assessment challenges of 360$\degree$ videos include  need to establish test procedures for subjects with different viewing device characteristics, lack of unified rating scale, limited well-defined objective metrics that can take into account the diverse quality of omnidirectional videos, difficulty in assessing quality degradation by taking into account eye movements across the viewport, different content characteristics, and impact of network impairments on quality ratings. Furthermore, the significant factors to consider are cybersickness and spatial presence, which arise from watching 360$\degree$ videos using VR headsets. These effects are exclusive to VR experiences and do not manifest when users watch normal videos on flat screens.\\
 2) \textit{Objective Quality Assessment:} Traditional objective VQA metrics are not suitable for evaluating the 360$\degree$ videos as the quality is significantly affected by network conditions, end-user viewing behavior, and current 2D planar projection methods that do not provide a uniform sample density at each pixel position. The quality metrics for 360$\degree$ videos involve remapping based on the relevant projection format. The experimental analysis demonstrates that the present objective metrics for 360$\degree$ videos (e.g., \cite{ssim_weighted}, \cite{psnr_vr}), when compared to the ground-truth quality
exhibit less correlation with the quality perceived subjectively.
Hence, there is a strong requirement for an optimal quality metric that can demonstrate a higher level of correlation with the ground truth, specifically tailored for 360$\degree$ content.
\\
3) \textit{Ensuring Visual Comfort:} The majority of the studies focus on utilizing an adaptive architecture determined by the viewport and distribution of quality throughout the entire image. Nevertheless, present research studies rarely prioritize the aspect of visual comfort. The user's enjoyment with 360$\degree $ video content is more affected by disturbance while using a headset compared to when using a traditional display. It is vital to lighten the weight of the head mounted device, making it more comfortable to wear. Additionally, decreasing cybersickness also plays a significant part in enhancing visual comfort. A number of viewers have claimed to experience fatigue, motion sickness, and nausea  while using VR. This has become one of the factors contributing to users' reluctance to embrace VR technology. The user experience must be prioritized while evaluating VR video streams. Research indicates that type of content (e.g., fast-moving, resolution), high usability, and presence ratings  can increase the likelihood of experiencing cybersickness. It is therefore imperative to investigate ways to lower the prevalence of cybersickness. Watching VR using HMDs adds factors like display configurations and rendering delays, which greatly affect QoE. Existing QoE models and conventional objective metrics, such as PSNR, fail to capture such factors. Therefore, it is crucial to create accurate mathematical models that are specifically designed for user characteristics in order to optimize individualized QoE frameworks. This area necessitates additional investigation in future VR studies. \\
 4) \textit{User interaction in AR:} When users interact with the real-world environment via AR technology, certain objective quality factors are taken into account. Some of these factors include the rate at which tasks are completed, the accuracy of interactions, and the time it takes for users to respond. User interaction (e.g., gesture, speech, and movement commands) is critical to the user's overall QoE.  Nevertheless, there isn't a reliable and standard metric to gauge how the users felt about the interaction. During the interaction, there is no way to get their feedback. Only \cite{ar_holo} used interaction in the objective QoE forecast.  Unfortunately, this problem has received minimal attention. Particularly in AR scenarios, there remain unresolved issues, such as the precise alignment of virtual objects with actual objects and the need to minimize both the rate of interaction errors and any delays that may occur.\\
   \textit{5)  QoE modeling and Complexity Analysis:} There is a lack of a comprehensive QoE model that accurately defines the effects of each influence factor in an adaptive streaming system. The discussed QoE predictor models can be  extended to include additional complex metrics and transmission/ streaming parameters as input features, but this can cause the complexity to increase. In comparison to the training set, the domain of potential inputs to the ML-based QoE models is much larger.
   It is still very difficult and mostly unsolved to find accurate and efficient VQA models that can make estimations for streaming services almost in real time.
   \par Additionally, the study of perceived quality in VR using different senses, like olfactory, tactile, and taste, is still extensively unexplored. Broadly, these sensations and emotions are difficult to quantify. There is a lack of a complete theoretical framework/model that considers and evaluates how people feel about these emotions. Precise evaluation of QoE is a necessary requirement for adaptive streaming solutions. It is a crucial element in maximizing the efficiency of 360$\degree$ video streaming service. So, there is a strong need for more systematic study in order to develop appropriate models and metrics for assessing the QoE in 360$\degree$ videos that are widely accepted.
  \par  Most of the studies lack discussion on the complexity of QoE model. Although demonstrating exceptional performance, the contemporary DL-based QoE models, necessitate substantial floating-point computations. Therefore, they are not practical or efficient to be widely used in their present state. Implementing complex QoE models can result in decreased application performance due to higher power consumption and greater use of computing resources.   Further, these models can be used to provide QoE feedback based solutions to achieve optimized resource allocation and efficient bandwidth utilization.  The end user is solely a consumer of the video service and lacks the ability to exert any influence, although there might be a possibility of including direct feedback in the future. However, client-side monitoring always  has risk of privacy invasion. Although a number of ML models have been used  for QoE evaluation and ML-based streaming techniques have been developed, the selection of the most appropriate ML model for a specific application is still a topic of ongoing research.\\
 6) \textit{Storage, transmission and rendering of VR content:} VR provides a more realistic viewing experience compared to the usual way of watching images and videos on phones, TVs, and computers with flat screens. The viewer is free to observe it in any direction due to the image's ability to occupy the full viewing space. At any instant, they can gaze at only a small part of the image. Their visual perception is determined by their visual attention, spatial arrangement of the image content, and the object that garners their attention.  Unrestricted access to high-resolution, immersive VR entails a substantial amount of data. This presents difficulties in terms of storage, transmission, and rendering of images, potentially impacting the quality of the watching experience. The VR applications require bigger file sizes, a variety of storage formats, and immersive viewing conditions.  This creates substantial challenges in obtaining, compressing, transmitting, and presenting high-quality VR video.\\
7) \textit{Open source tools and Applications:} Various open source tools such as Avtrack360 \cite{avtrack} and OpenTrack \cite{opentrack} are capable of capturing the head traces of users viewing 360$\degree$ videos, ALTRUIST \cite{altruist} can conduct subjective QoE assessment tests, Unity3D-based app \cite{unity3d} is available for 360$\degree$ VR quality measurement. There is still a need for a variety of tools and applications that can capture head traces, conduct VR subjective tests, and measure the quality of VR content. These should be made openly accessible to the public. The tools and apps must be user-friendly, and their accuracy must be ascertained.\\
8) \textit{Viewport Prediction:} There is a requirement to address additional concern associated with delivering immersive 360$\degree$ video content, such as ensuring correct prediction of the viewer's viewport over an extended period of time. Moreover, there is a perpetual query regarding the accuracy of viewport prediction. If video tiles are requested based on an incorrect estimate, the viewer's actual viewport may be obscured by black tiles for which no content has been requested. The viewer's movement exhibits significant volatility while watching certain portions of the video, which puts pressure on the training of ML models. The majority of viewport prediction approaches prioritize examining saliency patterns in addition to location information. In order to accurately forecast areas of attention and comprehend the correlation between the user's watching preference and saliency maps, it is necessary to enhance and properly train saliency models using extensive datasets, particularly those obtained from various camera rotations. However, since various motions may compromise the reliability of predictions, it is also necessary to examine the motion maps. There is a need to explore deep attention-based architectures and the involved computational complexity to improve the integration of multiple modalities (e.g., history of past viewings, video content) that vary in time and space. The viewport prediction models involve additional computational complexity,  and  most of the studies lack analysis on the computational complexity.\\
 9) \textit{Intelligent and adaptive multimedia streaming solutions:}  In a highly dynamic environment, the throughput estimation might not be very accurate. The proposed intelligent and adaptive video streaming algorithms do not consider errors caused by these incorrect estimates.
    All the heterogeneous aspects of the end-users like device characteristics (e.g., display resolution, battery constraints) can also be considered for more efficient user-centric streaming solutions.
     For efficient video streaming, the timely delivery of videos is of utmost importance, as it is significantly impacted by various factors such as dynamic network conditions and restricted transmission resources. Poor network conditions result in increased round-trip latency, which impacts the perceived quality~\cite{singhalVTC2023}.  
     Perceptual QoE driven resource allocation strategies can effectively meet the demands of video streaming users without compromising the viewing experience.
    There is often a lag between buffering and playback in the events of poor network condition. The effect on QoE due to the rebuffering events with the mentioned lag needs to be evaluated to study the extent of variation in lag between buffering and playback. Furthermore, adaptation strategies can be developed in the streaming session considering the lag factor.
   The different heterogeneity aspects of the end users (e.g., display resolution, battery constraints) should be considered for developing energy-efficient streaming solutions for diverse multimedia content and applications. Despite the emergence of several hybrid ML frameworks (such as integration of fuzzy logic with RL \cite{fred}) for intelligent and adaptive streaming, most studies \cite{ fred, robust_saliency, mosaic} lack performance evaluations on runtime analysis and GPU utilization. \\
10) \textit{Low latency multimedia streaming:} Multimedia streaming services, particularly 360$\degree$ and VR/AR videos, necessitate minimal response latency, encompassing network, buffering, edge processing, request overhead,  interaction, and feedback delay, all of which impact the video quality. With variation in user head movements, the brain anticipates immediate updates in auditory and visual content at different viewing angles. Consequently, ultra-low endpoint processing latency, minimal network latency, rapid edge computing are essential to achieve this degree of responsiveness to viewer head movements. In viewport based adaptive streaming, ultra-low network latency must be guaranteed to immediately provide the viewport due to the viewer's constant interaction via end-user devices. The caching and multicasting techniques mostly  focus on optimizing network-level parameters (latency) \cite{mahmoud_survey}, instead of optimizing the experienced perceptual quality. To mitigate latency in multi-user VR video streaming, it is essential to develop QoE-aware DL-assisted multicast framework. The next 5G and 6G networks are supposed to meet the substantial need for immersion, low latency, high capacity, highly reliable transmission, and real-time requirements in VR/AR application services \cite{metaverse_survey}.\\
 \textit{11) Efficient multimedia streaming solutions:} To enhance the streaming efficiency, it is viable to optimize the extent of viewport coverage and surrounding regions by dynamically adjusting them according to the user's head motions and prediction errors. In order to reduce the data rate requirement, upscaling the resolution of 360$\degree$ video segments on the client side from the downsampled ones result in loss of information that can impact the quality. The discussion persists on how to achieve a trade-off between minimizing information loss while simultaneously reducing the data rate. In tile-based streaming, tiles with varying quality based on user preferences are selected to maintain balance between quality and required bandwidth. Several factors influence this balance and further efforts are required to enhance QoE and minimize bandwidth usage. Enhanced design insights must be taken into account for the dynamic selection of tiles, with optimal allocation of bandwidth per tile within the framework of prioritized 360$\degree$ video delivery. Investigating the impact of variable size tiling on streaming performance and its impact on viewing experience with associated costs is another important issue. Priority based high resolution 360$\degree$ tiles, can be transmitted along the optimal available path in the event of multi-path transmission to enhance perceptual experience and transmission flexibility.
With a receiver-centric design, numerous users are anticipated to view several parts of the identical content. It is vital to enhance the resolution without increasing latency of the system. In order to achieve low latency, it is necessary to develop efficient compression techniques for 360$\degree$ data units, which is still an open challenge. New encoding techniques have to be developed to achieve better compression efficiency and have faster representation switches, hence providing reduced latency and computational expense.
\par 
\section{Conclusion }
\label{s:con}
This paper has reviewed the solutions pertinent to user-centric multimedia streaming, especially in modeling accurate video QoE predictors (overall as well as continuous and time varying), efficient (intelligent and adaptive) multimedia streaming using user-centric feedback, and intelligent 360$^\circ$ video streaming to reduce the data rate requirement while enhancing the immersive experience. It includes information on  how to prepare, process, and transmit the video content to end-user display devices, such as VR headsets, mobiles, smartphones, and monitors. Typically, the end users desire an acceptable QoE. Additionally, they seek an application that is user-friendly, without requiring manual adjustment prior to or during service usage. Other factors related to end users, such as the energy consumption of viewing devices and the amount of client bandwidth used, are also of relevance. We reviewed state-of-the-art ML-based QoE prediction models for video streaming and extended reality. In the context of ML applications, there has been a recent trend towards solutions utilizing LSTM for QoE evaluation and DRL for streaming frameworks. This tendency has been driven by the abundance of available data and the increased processing capability of compute devices. The QoE evaluation has witnessed a rise in the utilization of ML techniques due to their improved accuracy in predicting QoE. This also facilitates real-time, accurate, and flexible QoE management frameworks.
\par We have also discussed the various user-centric adaptive streaming techniques that consider inputs such as buffer level, throughput conditions, and chunk bit-rate. These schemes use ML approaches including the viewport and tile-based adaptation for 360$\degree$ videos. We have provided an overview of each scheme by outlining the issues they intend to address, their objectives, research findings, key elements, and major observation/ application.
The goal of these adaptive techniques is to ensure the highest possible perceptual quality while monitoring the resulting QoE at the client side that enables immediate feedback in order to enhance the QoE of a specific user. Researchers studying adaptive streaming may find our comparison useful since it provides a generic, consistent framework for explicitly evaluating, comparing, and testing the effectiveness of various bit-rate adaptation techniques.
Specifically, the requirements of 360$\degree$ videos are relatively different in terms of processing parameters (such as resolution, framerate, bit-rate, tile quality) as well as network transmission parameters (such as end-to-end latency, network capacity, low response delay) than conventional videos. Also from the users' perspective, the viewing behaviour and dynamic interaction with the scenes contribute to a distinctly unique immersive experience.  
\par Many research issues related to viewport prediction methods, tiling strategy, QoE evaluation, and the influence of additional limitations on 360$\degree$ video streaming to multiple VR clients is discussed.  The datasets are summarized that have records of throughput traces, video quality scores, head-movement, and eye-tracking logs which can be of further help to the researchers. We have listed several open challenges pertaining to the design of QoE predictor models and efficient multimedia  streaming techniques. The limitations with respect to the existing methods are summarized and a perspective is offered on possible solution approaches to address these research challenges, with an insight into possible directions where research may be extended. As a whole the review reflects on how the recent technologies have enhanced performance and overcome many problems with the aid of sophisticated and intelligent adaptive strategies. However, there still exists many open challenges in this domain.

{\footnotesize
\bibliography{mybibfile}
}
\vspace{5 cm}
\textbf{\Large{Glossary}}\\
\\
\textbf{ABR} Adaptive Bit-Rate.\\
\textbf{AR} Augmented Reality.\\
\textbf{AVC} Audio Video Coding.\\
\textbf{BB} Buffer Based.\\
\textbf{CMP} Cubic Mapping Projection.\\
\textbf{CNN} Convolutional Neural Network.\\
\textbf{DASH} Dynamic Adaptive Streaming over HTTP.\\
\textbf{DL} Deep Learning.\\
\textbf{DMOS} Differential Mean Opinion Score.\\
\textbf{DQN} Deep Q-Network.\\
\textbf{DRL} Deep Reinforcement Learning.\\
\textbf{ERP} Equirectangular Projection.\\
\textbf{FoV} Field of View.\\
\textbf{FR} Full Reference.\\
\textbf{GAN} Generative Adversarial Network.\\
\textbf{HD} High Definition.\\
\textbf{HEVC} High Efficiency Video Coding.\\
\textbf{HMD} Head-Mounted Display.\\
\textbf{IF} Influence Factor.\\
\textbf{IQA} Image Quality Assessment.\\
\textbf{ITU} International Telecommunication Union.\\
\textbf{k-NN} k-Nearest Neighbors.\\
\textbf{LSTM} Long Short-Term Memory.\\
\textbf{MCT} Motion Constrained Tiles.\\
\textbf{MDP} Markov Decision Process.\\
\textbf{ML} Machine Learning.\\
\textbf{MOS} Mean Opinion Score.\\
\textbf{MOVIE} MOtion-based Video Integrity Evaluator.\\
\textbf{MPD} Media Presentation Description.\\
\textbf{MS-SSIM} Multiscale SSIM.\\
\textbf{MSE} Mean Square Error.\\
\textbf{VP-NIQE} Visual Perception Natural Image Quality Evaluator.\\
\textbf{NR} No Reference.\\
\textbf{OR} Outlier Ratio.\\
\textbf{PLCC} Pearson Linear Correlation Coefficient.\\
\textbf{PSNR} Peak Signal to Noise Ratio.\\
\textbf{QoE} Quality of Experience.\\
\textbf{QoS} Quality of Service.\\
\textbf{QP} Quantization Parameter.\\
\textbf{RB} Rate Based.\\
\textbf{RL} Reinforcement Learning.\\
\textbf{RNN} Recurrent Neural Network.\\
\textbf{RR} Reduced Reference.\\
\textbf{SI} Spatial Information.\\
\textbf{SSIM} Structural Similarity Index.\\
\textbf{STSQ} Short Time Subjective Quality.\\
\textbf{SVC} Scalable Video Coding.\\
\textbf{SVM} Support Vector Machine.\\
\textbf{SVR} Support Vector Regression.\\
\textbf{TI} Temporal Information.\\
\textbf{TVSQ} Time Varying Subjective Quality.\\
\textbf{UHD} Ultra High Definition.\\
\textbf{VCEG} Video Coding Experts Group.\\
\textbf{VQA} Video Quality Assessment.\\
\textbf{VR} Virtual Reality.

\end{document}